\documentstyle[12pt]{article}

\begin{document}

\title{\Large {\bf Microcausality of Dirac field on noncommutative 
spacetime } }

\author{{{\large Zheng Ze Ma}} \thanks{Electronic address: 
           z.z.ma@seu.edu.cn} 
  \\  \\ {\normalsize {\sl Department of Physics, Southeast University, 
           Nanjing, 210096, P. R. China } } }

\date{}

\maketitle

\vskip 0.8cm

\centerline{\normalsize {\bf Abstract} ~~~ } 

\vskip 0.5cm 
 
\begin{minipage}{14.5cm}

\indent 

~~~~  We study the microcausality of free Dirac field on noncommutative 
spacetime. We calculate the vacuum and non-vacuum state expectation 
values for the Moyal commutator 
$[\overline{\psi}_{\alpha}(x)\star\psi_{\beta}(x),\overline{\psi}_
{\sigma}(x^{\prime})\star\psi_{\tau}(x^{\prime})]_{\star}$ of Dirac 
field on noncommutative spacetime. We find that they do not vanish 
for some cases of the indexes for an arbitrary spacelike interval, 
no matter whether $\theta^{0i}=0$ or $\theta^{0i}\neq0$. However for 
the physical observable quantities of Dirac field such as the Lorentz 
scalar $:\overline{\psi}(x)\star\psi(x):$ and the current 
$j^{\mu}(x)=:\overline{\psi}(x)\gamma^{\mu}\star\psi(x):$ etc., 
we find that they still satisfy the microcausality. Therefore 
microcausality is satisfied for Dirac field on 
noncommutative spacetime. 

\end{minipage}

\vskip 0.5cm 


\vskip 0.5cm

\section{Introduction}

\indent 

  In recent years, noncommutative field theories (NCFTs) have caused a 
lot of researches [1-5]. NCFTs have many different properties from 
that of the commutative spacetime field theories, such as the locality 
and unitarity. The spacetime noncommutativity is described by 
the commutation relations of the spacetime coordinates
$$
  [x^{\mu},x^{\nu}]=i\theta^{\mu\nu} ~,
  \eqno{(1.1)}  $$
where $\theta^{\mu\nu}$ is a constant real antisymmetric matrix with the 
dimension of square of length. For NCFTs, their Lagrangians can be obtained 
through replacing the ordinary products between field functions by the 
Moyal star-products. The Moyal star-product of two functions is given by 
\begin{eqnarray*}
  f(x)\star g(x) & = & e^{\frac{i}{2}\theta^{\mu\nu}\frac{\partial}{\partial 
                 \alpha^\mu}\frac{\partial}{\partial\beta^{\nu}}}f(x+\alpha)
                 g(x+\beta)\vert_{\alpha=\beta=0} \\
                 & = & f(x)g(x)+\sum\limits^{\infty}_{n=1}
                 \left(\frac{i}{2}\right)^{n}
                 \frac{1}{n!}\theta^{\mu_{1}\nu_{1}}
                 \cdots\theta^{\mu_{n}\nu_{n}}
                 \partial_{\mu_{1}}\cdots\partial_{\mu_{n}}f(x) 
                 \partial_{\nu_{1}}\cdots\partial_{\nu_{n}}g(x) ~.
\end{eqnarray*}
$$  \eqno{(1.2)}  $$

  In noncommutative spacetime, there may exist the interactions and 
waves with the propagation speed faster that the speed of light up to 
infinite [6-8]. This is because of the UV/IR mixing [9] and nonlinear 
interactions in NCFTs. We hope to study for the free fields on 
noncommutative 
spacetime whether there exist the information and interaction with 
the transmit speed faster than the speed of light. This can be acquainted 
from the microcausality of quantum fields on noncommutative spacetime. 
If the microcausality of free quantum fields on noncommutative spacetime 
is violated, then for the free fields on noncommutative spacetime, there 
exist the information and interaction with the transmit speed faster 
than the speed of light.

  According to superstring theories, people usually consider that the 
noncommutative parameters $\theta^{\mu\nu}$ are invariant constants. This 
will make the Lorentz invariance be violated for noncommutative field 
theories 
generally, except that a subgroup $SO(1,1)\times SO(2)$ of the usual 
Lorentz group can be maintained for certain special forms of the 
parameters $\theta^{\mu\nu}$ [10]. In Ref. [10], the authors constructed 
the $SO(1,1)\times SO(2)$ invariant spectral measure for the Fourier 
expansions of quantum fields on noncommutative spacetime. This makes the 
microcausality of quantum fields on noncommutative spacetime be 
formulated with respect to the $SO(1,1)$ light wedge. Because inside the 
light wedge, there are areas outside the light cone. Thus microcausality 
of quantum fields on noncommutative spacetime will be broken in the area 
of the light wedge outside the light cone. This results the existence 
of infinite propagation speed of free fields inside the light wedge. In 
Ref. [11], the authors pointed out that even the $SO(1,1)$ microcausality 
may be violated to consider the renormalization of the propagators. In 
Ref. [12], the authors generalized the axiom of Ref. [10] for the 
microcausality of quantum fields on noncommutative spacetime in 
accordance with the $SO(1,1)\times SO(2)$ invariance.

  It is necessary to point out that the infinite propagation speed of 
free fields on noncommutative spacetime of Ref. [10] is a necessary 
result of the breakdown of the Lorentz invariance, i.e., the breakdown 
of the usual Lorentz invariance of the spectral measure for the Fourier 
expansions of quantum fields. However there are doubts that whether 
the spectral measure for free fields on noncommutative spacetime is 
really in the form of $SO(1,1)\times SO(2)$ invariance as that 
constructed in Ref. [10]. 
Certes if the noncommutative parameters $\theta^{\mu\nu}$ are invariant 
constants, then Lorentz invariance of NCFTs will be destroyed [13,14]. 
However, there are possibilities that Lorentz invariance is maintained 
for NCFTs if we suppose that $\theta^{\mu\nu}$ carries tensor indexes. 
In Ref. [15], the authors suppose that $\theta^{\mu\nu}$ is a tensor 
operator and have constructed the Lorentz invariant NCFTs. 
In Refs. [16,17], the authors demonstrated the Lorentz invariance and 
unitarity for NCFTs to take $\theta^{\mu\nu}$ to be a tensor operator 
as that proposed in Ref. [15]. In Ref. [18], the authors demonstrated 
that Lorentz invariance 
can be maintained for the classical field equations of NCFTs if one 
take $\theta^{\mu\nu}$ to be a $c$-number tensor while not as an 
operator. In fact we can also demonstrate that Lorentz invariance 
is maintained for the $S$-matrixes of NCFTs if we take $\theta^{\mu\nu}$ 
to be a $c$-number tensor. On the other hand, the breakdown of Lorentz 
invariance has not been discovered yet in the experiments up to now [19]. 
It seems that Lorentz invariance is a more fundamental principle 
of physics, although it may not be so for a large scale structure 
of the universe, it should be satisfied in the local area of the 
universe. From the superstring theories, to take $\theta^{\mu\nu}$ 
to be a $c$-number tensor means that the background NS-NS ${\bf B}$-field 
changes as a second-order antisymmetric tensor when the reference system 
changes. On the other hand, it is necessary to point out that the 
nonlocality of NCFTs can be exist in accordance with the Lorentz 
invariance. The nonlocality of NCFTs does not mean that Lorentz 
invariance may be necessarily broken for NCFTs.

  Therefore we do not suppose that the spectral measure for the Fourier 
expansions of free fields on noncommutative spacetime is in the form 
of $SO(1,1)\times SO(2)$ invariance as that of Ref. [10]. As discovered 
in Ref. [20], in fact the light wedge can be resulted from the nonlocal 
interactions of NCFTs. For example for a six-dimensional theory, the 
usual Lorentz invariance can be reduced to $SO(3,1)\times SO(2)$ from 
the nonlocal interactions of NCFTs. Similarly, the light wedge of 
$SO(1,1)\times SO(2)$ can be resulted from the nonlocal interactions 
of a four-dimensional NCFT. Therefore we would rather think that the 
spectral measure of the $SO(1,1)\times SO(2)$ invariance for the Fourier 
expansions of free fields on noncommutative spacetime is an effective 
result that resulted from the nonlocal interactions of NCFTs.

  As pointed above, we hope to study whether there exist the information 
and interaction with the transmit speed faster than the speed of light 
for free fields on noncommutative spacetime through the microcausality 
property of quantum fields on noncommutative spacetime. We expand the free 
fields according to their usual form as that in the ordinary commutative 
spacetime. Another reason for us to take the Fourier expansion of quantum 
fields on noncommutative spacetime in their usual form is that in most 
occasions for the perturbative calculations of NCFTs in the literature, 
the propagators of quantum fields on noncommutative spacetime are obtained 
based on the usual Lorentz invariant spectral measures for the expansion 
of quantum fields. For the microcausality of the free scalar field on 
noncommutative spacetime, some 
results are obtained in Refs. [21-23]. In Ref. [22], Greenberg obtained 
that microcausality is violated for scalar field on noncommutative 
spacetime generally even if $\theta^{0i}=0$. However we have pointed out 
in Ref. [23] that there are some problems for the arguments of Ref. [22] 
for the result of the microcausality violation. In Ref. [23] we obtained 
that microcausality is violated for the quadratic operators of scalar 
field on noncommutative spacetime only when $\theta^{0i}\neq0$. In this 
paper, we will study the microcausality of free Dirac field on 
noncommutative spacetime.

  The spacetime noncommutative relations (1.1) is defined for 
coordinates of the same spacetime point. Because we will calculate 
the commutators of two operators on two different spacetime points 
in the following, we need to generalize the noncommutative 
relations (1.1) to two different spacetime points. Therefore 
we suppose  
$$
  [x_{1}^{\mu},x_{2}^{\nu}]=i\theta^{\mu\nu} ~. 
  \eqno{(1.3)}  $$
This make us possible define the Moyal star-product of two functions 
on two different spacetime points as [5]
\begin{eqnarray*}
  f(x_{1})\star g(x_{2}) & = & e^{\frac{i}{2}
          \theta^{\mu\nu}\frac{\partial}{\partial 
          \alpha^\mu}\frac{\partial}{\partial\beta^{\nu}}}f(x_{1}+\alpha)
                 g(x_{2}+\beta)\vert_{\alpha=\beta=0} \\
                 & = & f(x_{1})g(x_{2})+\sum\limits^{\infty}_{n=1}
                 \left(\frac{i}{2}\right)^{n}
                 \frac{1}{n!}\theta^{\mu_{1}\nu_{1}}
                 \cdots\theta^{\mu_{n}\nu_{n}}
                 \partial_{\mu_{1}}\cdots\partial_{\mu_{n}}f(x_{1}) 
                 \partial_{\nu_{1}}\cdots\partial_{\nu_{n}}g(x_{2}) ~. 
\end{eqnarray*}
$$  \eqno{(1.4)}  $$
A demonstration for the self-consistency of the commutative 
relations (1.3) with (1.1) is given in the Appendix.

  The content of this paper is organized as follows. In Sec. II, we 
analyze the criterion of microcausality violation for quantum fields 
on noncommutative spacetime. In Sec. III, we calculate the vacuum  
state expectation value for the Moyal commutator 
$[\overline{\psi}_{\alpha}(x)\star\psi_{\beta}(x),\overline{\psi}_
{\sigma}(x^{\prime})\star\psi_{\tau}(x^{\prime})]_{\star}$ of free 
Dirac field on noncommutative spacetime. We find that they do not 
vanish in some cases for an arbitrary spacelike interval, no matter 
whether $\theta^{0i}=0$ or $\theta^{0i}\neq0$. However for the 
physical observable quantities of Dirac field such as the Lorentz 
scalar $:\overline{\psi}(x)\star\psi(x):$ and the current 
$j^{\mu}(x)=:\overline{\psi}(x)\gamma^{\mu}\star\psi(x):$, we find 
that they still satisfy the microcausality. In Sec. IV, we generalize 
the result of Sec. III to the non-vacuum state expectation values and  
obtain that microcausality is satisfied for free Dirac field on 
noncommutative spacetime generally. Sec. V is the conclusion. In the 
Appendix, we give a demonstration for the self-consistency of the 
commutative relations (1.3) with (1.1).

\section{The criterion of microcausality violation} 

\indent 

  In this section, we first analyze the measurement of quantum fields 
on noncommutative spacetime and the criterion of microcausality 
violation. We need to point out that in the left hand sides of 
Eqs. (1.2) and (1.4), the coordinates $x^{\mu}$ are regarded as 
noncommutative 
operators that satisfying the commutation relations (1.1) and (1.3). 
In the right hand sides of Eqs. (1.2) and (1.4), after the expansion 
of Moyal star-products, the coordinates $x^{\mu}$ are treated as the 
ordinary commutative $c$-numbers, i.e., the experientially observable 
spacetime coordinates. To be clearer, we may write the coordinates in 
the left hand sides of Eqs. (1.2) and (1.4) as $\hat{x}^{\mu}$, in 
order to 
indicate that they are operators. However we do not use the sign 
$\hat{x}^{\mu}$ for the left hand sides of Eqs. (1.2) and (1.4), as 
well as in Eqs. (1.1) and (1.3). We do not discriminate them in signs 
and through out this paper. We consider that their prescribed meanings 
in different places are clear.

  For quantum field theories, as well as quantum mechanics, what the 
observer measures are certain expectation values. We suppose that there 
are two observers A and B situated at spacetime points $x$ and $y$, they 
proceed a measurement separately on the state vector $\vert\Psi\rangle$ 
for the locally observable quantity ${\cal O}(x)$ in the same occasion. 
However the time $x_{0}$ may not equal to the time $y_{0}$ generally. 
For the observer A, the state vector $\vert\Psi\rangle$ has been affected 
by the measurement of the observer B at the spacetime point $y$. Or we 
can say the observer B's observation instrument has taken an action on 
the state vector $\vert\Psi\rangle$. The state vector has become 
${\cal O}(y)\vert\Psi\rangle$. When the observer A takes his or her 
measurement on the state vector, his or her observation instrument 
will act on the state vector ${\cal O}(y)\vert\Psi\rangle$ again. 
These two sequent actions should be represented by the product 
operation of the operators. However because now the spacetime is 
noncommutative, the product operation should be the Moyal star-product, 
while not the ordinary product. Or we regard that in noncommutative 
spacetime, the basic product operation is the Moyal star-product. 
Thus what the measuring result the observer A obtained from his or her 
instrument is  
$\langle\Psi\vert{\cal O}(x)\star{\cal O}(y)\vert\Psi\rangle$. 
Similarly for the observer B, the state vector $\vert\Psi\rangle$ 
has been affected by the action of the observer A's instrument at 
the spacetime point $x$. The state vector becomes 
${\cal O}(x)\vert\Psi\rangle$. What the measuring result the 
observer B obtained from his or her instrument is 
$\langle\Psi\vert{\cal O}(y)\star{\cal O}(x)\vert\Psi\rangle$.

  Supposing that microcausality is satisfied for NCFTs, this means that 
there do not exist the 
physical information and interaction with the transmit speed faster 
than the speed of light. Thus when the spacetime interval between $x$ 
and $y$ is spacelike, the affection of the observer B's measurement 
or the action of observer B's instrument at spacetime point $y$ on 
the state vector $\vert\Psi\rangle$ has not propagated to the spacetime 
point $x$ when the observer A takes his or her measurement on the state 
vector $\vert\Psi\rangle$ at the spacetime point $x$. These two physical 
measurements do not interfere with each other. For the observer A, the 
state vector is still $\vert\Psi\rangle$, while not 
${\cal O}(y)\vert\Psi\rangle$. Therefore the measuring result what the 
observer A obtained is just 
$\langle\Psi\vert{\cal O}(x)\vert\Psi\rangle$. Thus we have 
$$
  \langle\Psi\vert{\cal O}(x)\star{\cal O}(y)]\vert\Psi\rangle=
     \langle\Psi\vert{\cal O}(x)\vert\Psi\rangle  
  ~~~~~~ \mbox{for} ~~~~~~ (x-y)^{2}<0 ~. 
  \eqno{(2.1)}  $$ 
The same reason as the observer A, the measuring result what the 
observer B obtained at the spacetime point $y$ is just 
$\langle\Psi\vert{\cal O}(y)\vert\Psi\rangle$. Thus we have 
$$
  \langle\Psi\vert{\cal O}(y)\star{\cal O}(x)]\vert\Psi\rangle=
     \langle\Psi\vert{\cal O}(y)\vert\Psi\rangle 
  ~~~~~~ \mbox{for} ~~~~~~ (x-y)^{2}<0 ~. 
  \eqno{(2.2)}  $$ 
We can suppose that the state vector $\vert\Psi\rangle$ is in the 
momentum eigenstate, thus it is also in the energy eigenstate. 
At the same time, we can suppose that the state vector 
$\vert\Psi\rangle$ is in the Heisenberg picture, therefore it does 
not rely on the spacetime coordinates. From the Heisenberg 
relations and the translation transformation, we have 
\footnote{
  We note $y-x=a$. From the Heisenberg relations and the translation 
  transformation, we have 
  $$  {\cal O}(y)=\exp(ia_{\mu}{\bf P}^{\mu}){\cal O}(x)
\exp(-ia_{\mu}{\bf P}^{\mu}) ~.                     $$  
Because $a_{\mu}$ now is a constant four-vector, from Eq. (1.2) we 
can also write the above expression as 
$$  {\cal O}(y)=\exp(ia_{\mu}{\bf P}^{\mu})\star{\cal O}(x)
\star\exp(-ia_{\mu}{\bf P}^{\mu}) ~.               $$  
  We use $P^{\mu}$ to represent the eigenvalues of the energy-momentum 
  of the state vector $\vert\Psi\rangle$. Therefore we obtain 
  \begin{eqnarray*}
& ~ & \langle\Psi\vert{\cal O}(y)\vert\Psi\rangle=
\langle\Psi\vert\exp(ia_{\mu}{\bf P}^{\mu})\star{\cal O}(x)
\star\exp(-ia_{\mu}{\bf P}^{\mu})\vert\Psi\rangle=
\langle\Psi\vert\exp(ia_{\mu}P^{\mu})\star{\cal O}(x)
\star\exp(-ia_{\mu}P^{\mu})\vert\Psi\rangle     \\
    & = & 
    \exp(ia_{\mu}P^{\mu})\star
    \langle\Psi\vert{\cal O}(x)\vert\Psi\rangle
\star\exp(-ia_{\mu}P^{\mu}) =
\exp(ia_{\mu}P^{\mu})
\langle\Psi\vert{\cal O}(x)\vert\Psi\rangle
\exp(-ia_{\mu}P^{\mu}) =           
    \langle\Psi\vert{\cal O}(x)\vert\Psi\rangle ~.
   \end{eqnarray*}
Therefore Eq. (2.3) is satisfied.}
$$
  \langle\Psi\vert{\cal O}(x)\vert\Psi\rangle=
       \langle\Psi\vert{\cal O}(y)\vert\Psi\rangle ~. 
  \eqno{(2.3)}  $$ 
Therefore the condition  
$$
  \langle\Psi\vert[{\cal O}(x),{\cal O}(y)]_{\star}\vert\Psi\rangle
    =\langle\Psi\vert{\cal O}(x)\star{\cal O}(y)\vert\Psi\rangle-
      \langle\Psi\vert{\cal O}(y)\star{\cal O}(x)\vert\Psi\rangle=0  
  ~~~~ \mbox{for} ~~~~ (x-y)^{2}<0  
  \eqno{(2.4)}  $$ 
should be satisfied for a NCFT to satisfy the microcausality.

  If microcausality is violated for a NCFT, 
then there may exist the physical information and interaction with 
the transmit speed faster than the speed of light. For the two 
measurements of the observer A and observer B located at $x$ and $y$ 
separated by a 
spacelike interval, the affection of the observer B's measurement at 
spacetime point $y$ on the state vector $\vert\Psi\rangle$ will 
propagate to the spacetime point $x$ when the observer A takes his 
or her measurement on the state vector $\vert\Psi\rangle$ at the 
spacetime point $x$, and the affection of the observer A's measurement 
at spacetime point $x$ on the state vector $\vert\Psi\rangle$ will 
propagate to the spacetime point $y$ when the observer B takes his 
or her measurement on the state vector $\vert\Psi\rangle$ at the 
spacetime point $y$. These two physical measurements will interfere 
with each other. For such a case, Eqs. (2.1) and (2.2) cannot be 
satisfied, while we still have 
$\langle\Psi\vert{\cal O}(x)\vert\Psi\rangle=
\langle\Psi\vert{\cal O}(y)\vert\Psi\rangle$ as that of Eq. (2.3). 
Therefore generally we have 
$$
  \langle\Psi\vert[{\cal O}(x),{\cal O}(y)]_{\star}
         \vert\Psi\rangle\neq0  
  ~~~~~~ \mbox{for} ~~~~~~ (x-y)^{2}<0 
  \eqno{(2.5)}  $$ 
for a NCFT to violate the microcausality. Therefore we can judge 
whether the microcausality is maintained or violated for a NCFT 
according to Eq. (2.5).

  Now we suppose that ${\cal O}_{1}(x)$ and ${\cal O}_{2}(y)$ are two 
different observable field operators, $x$ and $y$ are separated 
by a spacelike interval, two observers A and B situate at $x$ and $y$, 
and microcausality is satisfied for the field theory on noncommutative 
spacetime. Supposing that the observers A and B proceed a measurement 
separately on the state vector $\vert\Psi\rangle$ for the locally 
observable quantities ${\cal O}_{1}$ and ${\cal O}_{2}$ at $x$ and $y$ 
respectively, then from the above analysis, we have for the observer A 
$$
  \langle\Psi\vert{\cal O}_{1}(x)\star{\cal O}_{2}(y)]
\vert\Psi\rangle=\langle\Psi\vert{\cal O}_{1}(x)\vert\Psi\rangle  
  ~~~~~~ \mbox{for} ~~~~~~ (x-y)^{2}<0 ~.   
  \eqno{(2.6)}  $$ 
And we have for the observer B 
$$
  \langle\Psi\vert{\cal O}_{2}(y)\star{\cal O}_{1}(x)]
\vert\Psi\rangle=\langle\Psi\vert{\cal O}_{2}(y)\vert\Psi\rangle  
  ~~~~~~ \mbox{for} ~~~~~~ (x-y)^{2}<0 ~. 
  \eqno{(2.7)}  $$ 
Because now ${\cal O}_{1}(x)$ and ${\cal O}_{2}(y)$ are two 
different operators, we obtain generally 
$$
  \langle\Psi\vert{\cal O}_{1}(x)\vert\Psi\rangle\neq
       \langle\Psi\vert{\cal O}_{2}(y)\vert\Psi\rangle ~. 
  \eqno{(2.8)}  $$ 
Therefore from Eqs. (2.6)-(2.8) we have 
$$
  \langle\Psi\vert[{\cal O}_{1}(x),{\cal O}_{2}(y)]_{\star}\vert\Psi
\rangle\neq0 ~~~~~~ \mbox{for} ~~~~~~ (x-y)^{2}<0   
  \eqno{(2.9)}  $$
generally, even if $x$ and $y$ are separated by a spacelike interval, 
and the field theory satisfies the microcausality. Therefore we cannot 
deduce that a NCFT violates microcausality from Eq. (2.9) from the 
expectation values of the Moyal commutator of two different operators. 
In order to judge whether a NCFT violates microcausality, we must 
analyze the expectation values of the Moyal commutator of the same 
operator as that of Eq. (2.5).

\section{Vacuum state expectation values}

\indent 

  For the free Dirac field on noncommutative spacetime, its Lagrangian is 
given by 
$$
  {\cal L}=\overline{\psi}\star i\gamma^{\mu}\partial_{\mu}\psi-
            m\overline{\psi}\star\psi ~.
  \eqno{(3.1)}  $$
We suppose that in noncommutative spacetime, quantum fields can still be 
expanded in the usual Lorentz invariant form. Therefore the Fourier 
expansions for the free Dirac field and its conjugate field are given by 
$$
  \psi({\bf x},t)=\int\frac{d^{3}p}{(2\pi)^{3/2}}\sqrt
      {\frac{m}{E_{p}}}\sum\limits_{s=1,2}[b(p,s)u(p,s)
      e^{-ipx}+d^{\dagger}(p,s)v(p,s)e^{ipx}] ~,      $$
$$
  \overline{\psi}({\bf x},t)=\int\frac{d^{3}p}{(2\pi)^{3/2}}\sqrt
      {\frac{m}{E_{p}}}\sum\limits_{s=1,2}[b^{\dagger}(p,s)
       \overline{u}(p,s)e^{ipx}+d(p,s)\overline{v}(p,s)
       e^{-ipx}] ~,      
  \eqno{(3.2)}  $$ 
where $E_{p}=p_{0}=+\sqrt{\vert{\bf p}\vert^{2}+m^{2}}$ and 
$px=p_{\mu}x^{\mu}$. In Eq. (3.2), the spacetime coordinates are treated 
as noncommutative operators. They satisfy the commutation relations (1.1) 
and (1.3). The anticommutation relations for the creation and annihilation 
operators are still the same as that in the commutative spacetime. They 
are 
$$
  \{b(p,s),b^{\dagger}(p^{\prime},s^{\prime})\}=\delta_{ss^{\prime}}
           \delta^{3}({\bf p}-{\bf p}^{\prime}) ~,             $$     
$$
  \{d(p,s),d^{\dagger}(p^{\prime},s^{\prime})\}=\delta_{ss^{\prime}}
           \delta^{3}({\bf p}-{\bf p}^{\prime}) ~,             $$    
$$
  \{b(p,s),b(p^{\prime},s^{\prime})\}=
           \{d(p,s),d(p^{\prime},s^{\prime})\}=0 ~,            $$
$$
  \{b^{\dagger}(p,s),b^{\dagger}(p^{\prime},s^{\prime})\}=
       \{d^{\dagger}(p,s),d^{\dagger}(p^{\prime},s^{\prime})\}=0 ~,  $$
$$
  \{b(p,s),d(p^{\prime},s^{\prime})\}=
           \{b(p,s),d^{\dagger}(p^{\prime},s^{\prime})\}=0 ~,        $$
$$
  \{d(p,s),b(p^{\prime},s^{\prime})\}=
           \{d(p,s),b^{\dagger}(p^{\prime},s^{\prime})\}=0 ~.        
  \eqno{(3.3)}  $$
The spinors $u(p,s)$ and $v(p,s)$ satisfy the completeness relations
$$
  \sum\limits_{s=1,2}u_{\alpha}(p,s)\overline{u}_{\beta}(p,s)
         =\left(\frac{\not\!{p}+m}{2m}\right)_{\alpha\beta} ~,      
  ~~~~~~  
  \sum\limits_{s=1,2}v_{\alpha}(p,s)\overline{v}_{\beta}(p,s)
         =\left(\frac{\not\!{p}-m}{2m}\right)_{\alpha\beta} ~.      
  \eqno{(3.4)}  $$ 
They are the same as that of the commutative spacetime case.

  We define the Moyal anticommutators of the Dirac field to be 
$$
  \{\psi_{\alpha}(x),\overline{\psi}_{\beta}(x^{\prime})\}_{\star}
     =\psi_{\alpha}(x)\star\overline{\psi}_{\beta}(x^{\prime})+
      \overline{\psi}_{\beta}(x^{\prime})\star\psi_{\alpha}(x) ~,      
  \eqno{(3.5)}  $$
$$
  \{\psi_{\alpha}(x),\psi_{\beta}(x^{\prime})\}_{\star}
     =\psi_{\alpha}(x)\star\psi_{\beta}(x^{\prime})+
      \psi_{\beta}(x^{\prime})\star\psi_{\alpha}(x) ~,      
  \eqno{(3.6)}  $$
$$
  \{\overline{\psi}_{\alpha}(x),\overline{\psi}_{\beta}(x^{\prime})\}
_{\star}=\overline{\psi}_{\alpha}(x)\star\overline{\psi}_{\beta}
(x^{\prime})+\overline{\psi}_{\beta}(x^{\prime})\star
\overline{\psi}_{\alpha}(x) ~.      
  \eqno{(3.7)}  $$
We have shown in Ref. [24] that the Moyal anticommutators of the Dirac 
field are not $c$-number functions. In order to obtain the $c$-number 
results for these Moyal anticommutators, we need to evaluate their vacuum 
and non-vacuum state expectation values. We have obtained in Ref. [24] 
that 
$$
  \langle0\vert\{\psi_{\alpha}(x),\overline{\psi}_{\beta}
(x^{\prime})\}_{\star}\vert0\rangle
=\langle\Psi\vert\{\psi_{\alpha}(x),\overline{\psi}_{\beta}
(x^{\prime})\}_{\star}\vert\Psi\rangle  
  =-iS_{\alpha\beta}(x-x^{\prime}) ~,
  \eqno{(3.8)}  $$  
$$
  \langle0\vert\{\psi_{\alpha}(x),\psi_{\beta}
(x^{\prime})\}_{\star}\vert0\rangle = 
  \langle\Psi\vert\{\psi_{\alpha}(x),\psi_{\beta}
             (x^{\prime})\}_{\star}\vert\Psi\rangle=0 ~,          
  \eqno{(3.9)}  $$
$$
  \langle0\vert\{\overline{\psi}_{\alpha}(x),\overline{\psi}_{\beta}
             (x^{\prime})\}_{\star}\vert0\rangle = 
  \langle\Psi\vert\{\overline{\psi}_{\alpha}(x),\overline{\psi}_{\beta}
             (x^{\prime})\}_{\star}\vert\Psi\rangle =0 ~,              
  \eqno{(3.10)}  $$
where $\vert\Psi\rangle$ is a state vector of the Dirac field quantum 
system, and the singular function $S(x-x^{\prime})$ is given by [25]
$$
  S_{\alpha\beta}(x-x^{\prime})=
   -(i\not\!{\partial}_{x}+m)_{\alpha\beta}\Delta(x-x^{\prime}) ~.
  \eqno{(3.11)}  $$
$S(x-x^{\prime})$ is zero for a spacelike interval.

  For the free Dirac field on ordinary commutative spacetime, its 
observable quantities are constructed from the fundamental bilinear 
forms $\overline{\psi}_{\alpha}(x)\psi_{\beta}(x)$ and the 
$\gamma$-matrixes. For the fundamental bilinear forms 
$\overline{\psi}_{\alpha}(x)\psi_{\beta}(x)$, we have 
\begin{eqnarray*}
  & ~ & [\overline{\psi}_{\alpha}(x)\psi_{\beta}(x),
        \overline{\psi}_{\sigma}(x^{\prime})\psi_{\tau}
        (x^{\prime})]             \\ 
  & = & \overline{\psi}_{\alpha}(x)\{\psi_{\beta}(x),
        \overline{\psi}_{\sigma}(x^{\prime})\}
             \psi_{\tau}(x^{\prime})  
        -\{\overline{\psi}_{\alpha}(x),\overline{\psi}_{\sigma}
         (x^{\prime})\}
         \psi_{\beta}(x)\psi_{\tau}(x^{\prime})     \\
  & ~ & +\overline{\psi}_{\sigma}(x^{\prime})
         \overline{\psi}_{\alpha}(x)
         \{\psi_{\beta}(x),\psi_{\tau}(x^{\prime})\}
        -\overline{\psi}_{\sigma}(x^{\prime})
         \{\overline{\psi}_{\alpha}(x),
         \psi_{\tau}(x^{\prime})\}\psi_{\beta}(x) ~.
\end{eqnarray*} 
$$  \eqno{(3.12)}  $$
From the properties of the fundamental anticommutators of Dirac field, 
we have [25]
$$
  [\overline{\psi}_{\alpha}(x)\psi_{\beta}(x),
   \overline{\psi}_{\sigma}(x^{\prime})\psi_{\tau}(x^{\prime})]=0 
    ~~~~~~~~ \mbox{for} ~~~~~~~~ (x-x^{\prime})^{2}<0 ~. 
  \eqno{(3.13)}  $$
Therefore microcausality is satisfied for Dirac field on ordinary 
commutative spacetime.

  For the free Dirac field on noncommutative spacetime, its 
observable quantities such as the current 
$j^{\mu}(x)=:\overline{\psi}(x)\gamma^{\mu}\star\psi(x):$ are 
constructed from $:\overline{\psi}_{\alpha}(x)\star\psi_{\beta}(x):$ 
and $\gamma$-matrixes. Therefore in order to investigate its 
microcausality property, we need to analyze the Moyal anticommutator 
$[\overline{\psi}_{\alpha}(x)\star\psi_{\beta}(x),\overline{\psi}
_{\sigma}(x^{\prime})\star\psi_{\tau}(x^{\prime})]_{\star}$. We have 
\begin{eqnarray*}
  & ~ & [\overline{\psi}_{\alpha}(x)\star\psi_{\beta}(x),
        \overline{\psi}_{\sigma}(x^{\prime})\star\psi_{\tau}
        (x^{\prime})]_{\star} \\ 
  & = & \overline{\psi}_{\alpha}(x)\star\{\psi_{\beta}(x),
        \overline{\psi}_{\sigma}(x^{\prime})\}_{\star}
             \star\psi_{\tau}(x^{\prime})  
        -\{\overline{\psi}_{\alpha}(x),\overline{\psi}_{\sigma}
         (x^{\prime})\}_{\star}
         \star\psi_{\beta}(x)\star\psi_{\tau}(x^{\prime})     \\
  & ~ & +\overline{\psi}_{\sigma}(x^{\prime})\star
         \overline{\psi}_{\alpha}(x)
         \star\{\psi_{\beta}(x),\psi_{\tau}(x^{\prime})\}_{\star}
        -\overline{\psi}_{\sigma}(x^{\prime})\star
         \{\overline{\psi}_{\alpha}(x),
         \psi_{\tau}(x^{\prime})\}_{\star}\star\psi_{\beta}(x)  \\
  & = & \overline{\psi}_{\alpha}(x)\star\psi_{\beta}(x)\star 
        \overline{\psi}_{\sigma}(x^{\prime})\star\psi_{\tau}
        (x^{\prime})-\overline{\psi}_{\sigma}(x^{\prime})\star
        \psi_{\tau}(x^{\prime})\star
        \overline{\psi}_{\alpha}(x)\star\psi_{\beta}(x) ~.
\end{eqnarray*} 
$$  \eqno{(3.14)}  $$
As obtained in Ref. [24], the fundamental Moyal anticommutators of 
the Dirac field are not $c$-number functions, in order to examine the 
microcausality property for the operator 
$\overline{\psi}_{\alpha}(x)\star\psi_{\beta}(x)$, we need to 
calculate its expectation values, to see whether they are vanished 
or not for a spacelike interval. As demonstrated in Sec. II, this is 
also the demand of physical measurements. In order to simplify the 
calculation, we can adopt the normal orderings for the Moyal 
star-product. Therefore we need to calculate the function 
$$
  B_{\alpha\beta\sigma\tau}(x,y)=\langle\Psi\vert
[:\overline{\psi}_{\alpha}(x)\star
      \psi_{\beta}(x):,:\overline{\psi}_{\sigma}(y)\star
      \psi_{\tau}(y):]_{\star}\vert\Psi\rangle ~, 
  \eqno{(3.15)}  $$
where $\vert\Psi\rangle$ is a state vector of free Dirac field quantum 
system. To adopt the normal orderings for the field operators 
$\overline{\psi}_{\alpha}(x)\star\psi_{\beta}(x)$ and 
$\overline{\psi}_{\sigma}(y)\star\psi_{\tau}(y)$ means that an infinite 
charge of the vacuum with all of the negative energy states occupied has 
been eliminated in the corresponding commutative spacetime field theory. 
To be the limit of physical measurements, we take the state vector 
$\vert\Psi\rangle$ in Eq. (3.15) to be the vacuum state $\vert0\rangle$. 
Therefore in this section, we first study the vacuum expectation value 
$$
  B_{0,\alpha\beta\sigma\tau}(x,y)
=\langle0\vert[:\overline{\psi}_{\alpha}(x)\star
      \psi_{\beta}(x):,:\overline{\psi}_{\sigma}(y)\star
      \psi_{\tau}(y):]_{\star}\vert0\rangle ~. 
  \eqno{(3.16)}  $$
For the non-vacuum state expectation value of Eq. (3.15), we will 
analyze it in Sec. IV.

  We decompose $\psi(x)$ into the annihilation (positive frequency) 
and creation (negative frequency) part 
$$
  \psi(x)=\psi^{+}(x)+\psi^{-}(x) ~,                
  \eqno{(3.17)}  $$
where 
$$
  \psi^{+}(x)=\int\frac{d^{3}p}{(2\pi)^{3/2}}\sqrt{\frac{m}{E_{p}}}
      \sum\limits_{s=1,2}b(p,s)u(p,s)e^{-ipx} ~,              $$
$$
  \psi^{-}(x)=\int\frac{d^{3}p}{(2\pi)^{3/2}}\sqrt
      {\frac{m}{E_{p}}}\sum\limits_{s=1,2}d^{\dagger}(p,s)v(p,s)
        e^{ipx} ~.      
  \eqno{(3.18)}  $$
For the conjugate field we have 
$$
  \overline\psi(x)=\overline\psi^{+}(x)+\overline\psi^{-}(x) ~, 
  \eqno{(3.19)}  $$
$$
  \overline\psi^{+}(x)=\int\frac{d^{3}p}{(2\pi)^{3/2}}\sqrt
      {\frac{m}{E_{p}}}\sum\limits_{s=1,2}d(p,s)\overline{v}(p,s)
       e^{-ipx} ~,          $$
$$
  \overline\psi^{-}(x)=\int\frac{d^{3}p}{(2\pi)^{3/2}}\sqrt
      {\frac{m}{E_{p}}}\sum\limits_{s=1,2}b^{\dagger}(p,s)
       \overline{u}(p,s)e^{ipx} ~.         
  \eqno{(3.20)}  $$
From Eqs. (3.17) and (3.19) we have 
$$
  \overline{\psi}_{\alpha}(x)\star\psi_{\beta}(x)
   =\overline{\psi}_{\alpha}^{+}(x)\star\psi_{\beta}^{+}(x)+
    \overline{\psi}_{\alpha}^{+}(x)\star\psi_{\beta}^{-}(x)+
    \overline{\psi}_{\alpha}^{-}(x)\star\psi_{\beta}^{+}(x)+
    \overline{\psi}_{\alpha}^{-}(x)\star\psi_{\beta}^{-}(x) ~.
  \eqno{(3.21)}  $$ 
The normal ordering of the operator 
$\overline{\psi}_{\alpha}(x)\star\psi_{\beta}(x)$ is given by 
$$
  :\overline{\psi}_{\alpha}(x)\star\psi_{\beta}(x):
   =\overline{\psi}_{\alpha}^{+}(x)\star\psi_{\beta}^{+}(x)-
    \psi_{\beta}^{-}(x)\star\overline{\psi}_{\alpha}^{+}(x)+
    \overline{\psi}_{\alpha}^{-}(x)\star\psi_{\beta}^{+}(x)+
    \overline{\psi}_{\alpha}^{-}(x)\star\psi_{\beta}^{-}(x) ~.
  \eqno{(3.22)}  $$ 
Here we have made a simplified manipulation for the normal ordering 
of the Moyal star-product operator 
$\overline{\psi}_{\alpha}^{+}(x)\star\psi_{\beta}^{-}(x)$. 
This is because the result of the Moyal star-product of two 
functions is related with the order of two functions. In the 
Fourier integral representation, we can see that 
$\psi_{\beta}^{-}(x)\star\overline{\psi}_{\alpha}^{+}(x)$ 
will have an additional phase factor $e^{ip\times p^{\prime}}$ relative 
to $\overline{\psi}_{\alpha}^{+}(x)\star\psi_{\beta}^{-}(x)$. 
However in Eq. (3.22) we have ignored such a difference for the normal 
ordering of $\overline{\psi}_{\alpha}^{+}(x)\star\psi_{\beta}^{-}(x)$. 
The reason is that the terms that contain 
$\psi_{\beta}^{-}(x)\star\overline{\psi}_{\alpha}^{+}(x)$ in 
the expansion of Eq. (3.16) will contribute zero when we evaluate their  
vacuum expectation values, as it can be seen in the following. Thus we 
can ignore such a difference equivalently for convenience.

  To expand $:\overline{\psi}_{\alpha}(x)\star\psi_{\beta}(x):\star
:\overline{\psi}_{\sigma}(y)\star\psi_{\tau}(y):$, we obtain 
\begin{eqnarray*}
  & ~ & :\overline{\psi}_{\alpha}(x)\star\psi_{\beta}(x):\star
:\overline{\psi}_{\sigma}(y)\star\psi_{\tau}(y):         \\ 
  & = & \overline{\psi}_{\alpha}^{+}(x)\star\psi_{\beta}^{+}(x)\star
        \overline{\psi}_{\sigma}^{+}(y)\star\psi_{\tau}^{+}(y) -   
        \overline{\psi}_{\alpha}^{+}(x)\star\psi_{\beta}^{+}(x)\star 
        \psi_{\tau}^{-}(y)\star\overline{\psi}_{\sigma}^{+}(y)   \\
  & ~ & +\overline{\psi}_{\alpha}^{+}(x)\star\psi_{\beta}^{+}(x)\star
         \overline{\psi}_{\sigma}^{-}(y)\star\psi_{\tau}^{+}(y)
        +\overline{\psi}_{\alpha}^{+}(x)\star\psi_{\beta}^{+}(x)\star
         \overline{\psi}_{\sigma}^{-}(y)\star\psi_{\tau}^{-}(y)   \\
& ~ & -\psi_{\beta}^{-}(x)\star\overline{\psi}_{\alpha}^{+}(x)\star
         \overline{\psi}_{\sigma}^{+}(y)\star\psi_{\tau}^{+}(y) 
        +\psi_{\beta}^{-}(x)\star\overline{\psi}_{\alpha}^{+}(x)\star 
         \psi_{\tau}^{-}(y)\star\overline{\psi}_{\sigma}^{+}(y)   \\
& ~ & -\psi_{\beta}^{-}(x)\star\overline{\psi}_{\alpha}^{+}(x)\star
         \overline{\psi}_{\sigma}^{-}(y)\star\psi_{\tau}^{+}(y)
      -\psi_{\beta}^{-}(x)\star\overline{\psi}_{\alpha}^{+}(x)\star
         \overline{\psi}_{\sigma}^{-}(y)\star\psi_{\tau}^{-}(y)   \\
& ~ & +\overline{\psi}_{\alpha}^{-}(x)\star\psi_{\beta}^{+}(x)\star
        \overline{\psi}_{\sigma}^{+}(y)\star\psi_{\tau}^{+}(y)   
        -\overline{\psi}_{\alpha}^{-}(x)\star\psi_{\beta}^{+}(x)\star 
        \psi_{\tau}^{-}(y)\star\overline{\psi}_{\sigma}^{+}(y)    \\
  & ~ & +\overline{\psi}_{\alpha}^{-}(x)\star\psi_{\beta}^{+}(x)\star
         \overline{\psi}_{\sigma}^{-}(y)\star\psi_{\tau}^{+}(y)
        +\overline{\psi}_{\alpha}^{-}(x)\star\psi_{\beta}^{+}(x)\star
         \overline{\psi}_{\sigma}^{-}(y)\star\psi_{\tau}^{-}(y)   \\
& ~ & +\overline{\psi}_{\alpha}^{-}(x)\star\psi_{\beta}^{-}(x)\star
        \overline{\psi}_{\sigma}^{+}(y)\star\psi_{\tau}^{+}(y)    
        -\overline{\psi}_{\alpha}^{-}(x)\star\psi_{\beta}^{-}(x)\star 
        \psi_{\tau}^{-}(y)\star\overline{\psi}_{\sigma}^{+}(y)   \\
  & ~ & +\overline{\psi}_{\alpha}^{-}(x)\star\psi_{\beta}^{-}(x)\star
         \overline{\psi}_{\sigma}^{-}(y)\star\psi_{\tau}^{+}(y)
        +\overline{\psi}_{\alpha}^{-}(x)\star\psi_{\beta}^{-}(x)\star
         \overline{\psi}_{\sigma}^{-}(y)\star\psi_{\tau}^{-}(y) ~.
\end{eqnarray*}
$$  \eqno{(3.23)}  $$ 
From Eq. (3.23), we can clearly see that the non-zero contributions  
to the vacuum expectation value of 
$:\overline{\psi}_{\alpha}(x)\star\psi_{\beta}(x):\star
:\overline{\psi}_{\sigma}(y)\star\psi_{\tau}(y):$ come 
from the terms which the most right hand side components of the 
product operators are the negative frequency operators, and at the 
same time for these terms the number of the positive frequency 
component operators are equal to the number of the negative frequency 
component operators in the total product operators. Therefore we can 
see that there is only one such term 
$\overline{\psi}_{\alpha}^{+}(x)\star\psi_{\beta}^{+}(x) 
 \star\overline{\psi}_{\sigma}^{-}(y)\star\psi_{\tau}^{-}(y)$ 
which will contribute to the non-zero vacuum expectation value. 
Hence we have 
$$
  \langle0\vert:\overline{\psi}_{\alpha}(x)\star\psi_{\beta}(x):
    \star:\overline{\psi}_{\sigma}(y)\star\psi_{\tau}(y):
    \vert0\rangle = \langle0\vert\overline{\psi}_{\alpha}^{+}(x)
    \star\psi_{\beta}^{+}(x)
    \star\overline{\psi}_{\sigma}^{-}(y)\star\psi_{\tau}^{-}(y)
    \vert0\rangle ~. 
  \eqno{(3.24)}  $$ 
Similarly, the non-zero contribution to the vacuum expectation 
value of $\overline{\psi}_{\sigma}(y)\star\psi_{\tau}(y):
\star:\overline{\psi}_{\alpha}(x)\star\psi_{\beta}(x):$ only comes 
from the part 
$\overline{\psi}_{\sigma}^{+}(y)\star\psi_{\tau}^{+}(y)
\star\overline{\psi}_{\alpha}^{-}(x)\star\psi_{\beta}^{-}(x)$. 
We have 
$$
  \langle0\vert:\overline{\psi}_{\sigma}(y)\star\psi_{\tau}(y):
    \star:\overline{\psi}_{\alpha}(x)\star\psi_{\beta}(x):
    \vert0\rangle = \langle0\vert\overline{\psi}_{\sigma}^{+}(y)
    \star\psi_{\tau}^{+}(y)
    \star\overline{\psi}_{\alpha}^{-}(x)\star\psi_{\beta}^{-}(x)
    \vert0\rangle ~. 
  \eqno{(3.25)}  $$ 
If we do not use the normal orderings for the operators 
$:\overline{\psi}_{\alpha}(x)\star\psi_{\beta}(x):$ and 
$:\overline{\psi}_{\sigma}(y)\star\psi_{\tau}(y):$ as that 
of Eq. (3.14), then in the calculation of the vacuum expectation 
value for Eq. (3.14), we need to consider the additional four terms 
which will contribute non-zero results 
$$
  \overline{\psi}_{\alpha}^{+}(x)\star\psi_{\beta}^{-}(x)\star
  \overline{\psi}_{\sigma}^{+}(y)\star\psi_{\tau}^{-}(y)-
\overline{\psi}_{\sigma}^{+}(y)\star\psi_{\tau}^{-}(y)\star
  \overline{\psi}_{\alpha}^{+}(x)\star\psi_{\beta}^{-}(x) ~,          $$
$$
  \overline{\psi}_{\alpha}^{-}(x)\star\psi_{\beta}^{+}(x)\star
  \overline{\psi}_{\sigma}^{+}(y)\star\psi_{\tau}^{-}(y)-
\overline{\psi}_{\sigma}^{-}(y)\star\psi_{\tau}^{+}(y)\star
  \overline{\psi}_{\alpha}^{+}(x)\star\psi_{\beta}^{-}(x) ~.          $$
However in fact we can obtain that the total vacuum expectation value 
of these terms cancel at last. Thus to take the normal orderings 
for the operators $:\overline{\psi}_{\alpha}(x)\star\psi_{\beta}(x):$ 
and $:\overline{\psi}_{\sigma}(y)\star\psi_{\tau}(y):$ has simplified 
the calculation.

  Through calculation we obtain 
\begin{eqnarray*}
  & ~ & \langle0\vert[:\overline{\psi}_{\alpha}(x)\star
      \psi_{\beta}(x):,:\overline{\psi}_{\sigma}(y)\star
      \psi_{\tau}(y):]_{\star}\vert0\rangle             \\
  & = & \langle0\vert\overline{\psi}_{\alpha}^{+}(x)
    \star\psi_{\beta}^{+}(x)
    \star\overline{\psi}_{\sigma}^{-}(y)\star\psi_{\tau}^{-}(y)
    \vert0\rangle - \langle0\vert\overline{\psi}_{\sigma}^{+}(y)
    \star\psi_{\tau}^{+}(y)
    \star\overline{\psi}_{\alpha}^{-}(x)\star\psi_{\beta}^{-}(x)
    \vert0\rangle                      \\
  & = & \int\frac{d^{3}p_{1}}{(2\pi)^{3}}
        \int\frac{d^{3}p_{2}}{(2\pi)^{3}}
        \sum\limits_{s_{1},s_{2}}\Bigg[\left(\frac{m}{E_{p_{1}}}
        \right)^{\frac{1}{2}}\overline{v}_{\alpha}(p_{1},s_{1})
         e^{-ip_{1}x}\star
         \left(\frac{m}{E_{p_{2}}}\right)^{\frac{1}{2}}
         u_{\beta}(p_{2},s_{2})e^{-ip_{2}x}       \\
  & ~ & ~~~~~~~~~~~~~
          \star\left(\frac{m}{E_{p_{2}}}
        \right)^{\frac{1}{2}}\overline{u}_{\sigma}(p_{2},s_{2})
         e^{ip_{2}y}
        \star\left(\frac{m}{E_{p_{1}}}\right)^{\frac{1}{2}}
         v_{\tau}(p_{1},s_{1})e^{ip_{1}y}-          \\
  & ~ & ~~~~~~~~~~~~~~~~~~~~~~~~~~~~~~
          \left(\frac{m}{E_{p_{1}}}
        \right)^{\frac{1}{2}}\overline{v}_{\sigma}(p_{1},s_{1})
         e^{-ip_{1}y}
        \star\left(\frac{m}{E_{p_{2}}}\right)^{\frac{1}{2}}
         u_{\tau}(p_{2},s_{2})e^{-ip_{2}y}       \\ 
  & ~ & ~~~~~~~~~~~~~
          \star\left(\frac{m}{E_{p_{2}}}
        \right)^{\frac{1}{2}}\overline{u}_{\alpha}(p_{2},s_{2})
         e^{ip_{2}x}
        \star\left(\frac{m}{E_{p_{1}}}\right)^{\frac{1}{2}}
         v_{\beta}(p_{1},s_{1})e^{ip_{1}x}\Bigg]      \\
  & = & \int\frac{d^{3}p_{1}}{(2\pi)^{3}2E_{p_{1}}}
        \int\frac{d^{3}p_{2}}{(2\pi)^{3}2E_{p_{2}}}
        \Big[(\not\!{p}_{1}-m)_{\tau\alpha }
(\not\!{p}_{2}+m)_{\beta\sigma}
e^{-i(p_{1}+p_{2})(x-y)}                \\
  & ~ & ~~~~~~~~~~~~~~~~~~~~~~~~~~~~~~~~~~~~~~~
-(\not\!{p}_{1}-m)_{ \beta\sigma}
(\not\!{p}_{2}+m)_{\tau\alpha}
  e^{i(p_{1}+p_{2})(x-y)}\Big] ~.  
\end{eqnarray*}
$$  \eqno{(3.26)}  $$ 
In Eq. (3.26), there are two terms in the second equality. The 
first term means that two Dirac field quanta 
$\vert p_{1},s_{1}\rangle$ and $\vert p_{2},s_{2}\rangle$ are 
generated at the spacetime point $y$, and annihilated at the 
spacetime point $x$. The second term means that two Dirac field 
quanta $\vert p_{1},s_{1}\rangle$ and $\vert p_{2},s_{2}\rangle$ 
are generated at the spacetime point $x$, and annihilated at the 
spacetime point $y$. 
In Eq. (3.26), $\int\frac{d^{3}p_{1}}{(2\pi)^{3}2E_{p_{1}}}
\int\frac{d^{3}p_{2}}{(2\pi)^{3}2E_{p_{2}}}$ is the Lorentz 
invariant volume element, $\not\!\!{p}_{1}=p_{1\mu}\gamma^{\mu}$, 
$\not\!{p}_{2}=p_{2\mu}\gamma^{\mu}$, 
and $(p_{1}+p_{2})(x-y)=(p_{1}+p_{2})_{\mu}(x-y)^{\mu}$. Thus the 
total expression is Lorentz invariant. In the above calculation, 
we have used Eq. (1.4) for the Moyal star-product of two functions 
defined on two different spacetime points. The result does not 
rely on the parameters $\theta^{\mu\nu}$. However if 
$\theta^{\mu\nu}=0$, we can deduce from Eq. (3.13) directly that 
the free Dirac field satisfies the microcausality on ordinary 
commutative spacetime. For such a case, we need not to evaluate 
the expectation value of Eq. (3.26).

  We need to analyze whether the expression of Eq. (3.26) disappears 
or not for a spacelike interval. This can be seen from the vacuum 
expectation value of the equal-time commutator. Thus to take 
$x_{0}=y_{0}$ in Eq. (3.26), we have 
\begin{eqnarray*}
  & ~ & \langle0\vert[:\overline{\psi}_{\alpha}({\bf x},t)\star
      \psi_{\beta}({\bf x},t):,:\overline{\psi}_{\sigma}
       ({\bf y},t)\star\psi_{\tau}({\bf y},t):]_{\star}
        \vert0\rangle             \\
  & = & \int\frac{d^{3}p_{1}}{(2\pi)^{3}2E_{p_{1}}}
        \int\frac{d^{3}p_{2}}{(2\pi)^{3}2E_{p_{2}}}
        \Big[(\not\!{p}_{1}-m)_{\tau\alpha }
(\not\!{p}_{2}+m)_{\beta\sigma}
         e^{i({\bf p}_{1}+{\bf p}_{2})\cdot({\bf x}-{\bf y})}  \\
  & ~ & ~~~~~~~~~~~~~~~~~~~~~~~~~~~~~~~~~~~~~~~~~
- (\not\!{p}_{1}-m)_{ \beta\sigma}
(\not\!{p}_{2}+m)_{\tau\alpha}
         e^{-i({\bf p}_{1}+{\bf p}_{2})\cdot({\bf x}-{\bf y})}\Big] ~.
\end{eqnarray*}
$$  \eqno{(3.27)}  $$ 
We can see that in Eq. (3.27), the integral measure does not change 
when the arguments (${\bf p}_{1}$,${\bf p}_{2}$) change to 
($-{\bf p}_{1}$,$-{\bf p}_{2}$). The integral space is symmetrical 
to the integral arguments (${\bf p}_{1}$,${\bf p}_{2}$) and 
($-{\bf p}_{1}$,$-{\bf p}_{2}$). Therefore the odd function part 
of the integrand contributes zero to the whole integral. While the 
even function part will contribute nonzero to the whole integral. 
Thus to omit the odd function part in the integrand we obtain 
\begin{eqnarray*}
  & ~ & \langle0\vert[:\overline{\psi}_{\alpha}({\bf x},t)\star
      \psi_{\beta}({\bf x},t):,:\overline{\psi}_{\sigma}
       ({\bf y},t)\star\psi_{\tau}({\bf y},t):]_{\star}
        \vert0\rangle             \\
  & = & \int\frac{d^{3}p_{1}}{(2\pi)^{3}2E_{p_{1}}}
        \int\frac{d^{3}p_{2}}{(2\pi)^{3}2E_{p_{2}}}
         \Big[(mp_{10}\gamma^{0}_{\tau\alpha}\delta_{\beta\sigma} 
-mp_{10}\gamma^{0}_{\beta\sigma}\delta_{\tau\alpha}    
+mp_{20}\gamma^{0}_{\tau\alpha}\delta_{\beta\sigma}
-mp_{20}\gamma^{0}_{\beta\sigma}\delta_{\tau\alpha}  \\
  & ~ & ~~~~~~~~~~~~~~~~~~~~~~~~~~~~~~~~~~~~
+p_{1i}p_{2j}\gamma^{i}_{\tau\alpha}\gamma^{j}_{\beta\sigma} 
        -p_{1i}p_{2j}\gamma^{i}_{\beta\sigma}\gamma^{j}_{\tau\alpha}) 
\cos({\bf p}_{1}+{\bf p}_{2})\cdot({\bf x}-{\bf y})     \\
  & ~ & ~
      -(p_{10}p_{2j}\gamma^{0}_{\tau\alpha}\gamma^{j}_{\beta\sigma}
      +p_{1i}p_{20}\gamma^{i}_{\tau\alpha}\gamma^{0}_{\beta\sigma}
      +mp_{1i}\gamma^{i}_{\tau\alpha}\delta_{\beta\sigma}
-mp_{2j}\delta_{\tau\alpha}\gamma^{j}_{\beta\sigma}         \\
  & ~ & ~~~
+p_{10}p_{2j}\gamma^{0}_{\beta\sigma}\gamma^{j}_{\tau\alpha} 
+p_{1i}p_{20}\gamma^{i}_{\beta\sigma}\gamma^{0}_{\tau\alpha} 
      +mp_{1i}\gamma^{i}_{\beta\sigma}\delta_{\tau\alpha} 
-mp_{2j}\delta_{\beta\sigma}\gamma^{j}_{\tau\alpha})
~ i\sin({\bf p}_{1}+{\bf p}_{2})\cdot({\bf x}-{\bf y})\Big] ~, 
\end{eqnarray*}
$$  \eqno{(3.28)}  $$ 
where $p_{1i}=(p_{1x},p_{1y},p_{1z})$, $p_{2i}=(p_{2x},p_{2y},p_{2z})$. 
We can see that Eq. (3.28) does not vanish generally for an arbitrary 
interval of $({\bf x}-{\bf y})$ for some cases of the indexes $\alpha$, 
$\beta$, $\sigma$, and $\tau$, and the result depend on the choice of 
the representation of $\gamma$-matrixes. Because the total expression 
of Eq. (3.26) is Lorentz invariant, we have 
$$
  \langle0\vert[:\overline{\psi}_{\alpha}(x)\star
      \psi_{\beta}(x):,:\overline{\psi}_{\sigma}(y)\star
      \psi_{\tau}(y):]_{\star}\vert0\rangle\neq0 
  ~~~~~~~~ \mbox{for} ~~~~~~~~ (x-y)^{2}<0  
  \eqno{(3.29)}  $$ 
for some cases of the indexes $\alpha$, $\beta$, $\sigma$, and 
$\tau$, and the result depend on the choice of the representation 
of $\gamma$-matrixes. The result does not depend on whether 
$\theta^{0i}$ vanishes or not.

  However we cannot conclude that microcausality is violated 
necessarily for free Dirac field on noncommutative spacetime from the
above result. There are two sides of the reasons. On the one side as 
pointed out in Sec. II, for the criterion of the violation of 
microcausality, we must analyze the commutator of the same operator. 
Therefore in Eqs. (3.26)-(3.28), we must let $\sigma=\alpha$ and 
$\tau=\beta$. For such a case, we can see that in Eq. (3.28), the 
coefficient of $\cos({\bf p}_{1}+{\bf p}_{2})\cdot({\bf x}-{\bf y})$ 
is zero, however the coefficient of 
$\sin({\bf p}_{1}+{\bf p}_{2})\cdot({\bf x}-{\bf y})$ may not be zero 
for an arbitrary choice of the representation of $\gamma$-matrixes. 
Therefore we have 
$$
  B_{0,\alpha\beta\alpha\beta}(x,y)
=\langle0\vert[:\overline{\psi}_{\alpha}(x)\star
      \psi_{\beta}(x):,:\overline{\psi}_{\alpha}(y)\star
      \psi_{\beta}(y):]_{\star}\vert0\rangle\neq 0 
  ~~~~~~ \mbox{for} ~~~~~~ (x-y)^{2}<0  
  \eqno{(3.30)}  $$ 
for an arbitrary choice of the representation of $\gamma$-matrixes.

  If for Dirac field, the observable quantities are directly 
$\overline{\psi}_{\alpha}(x)\star\psi_{\beta}(x)$, then from 
Eq. (3.30) we can obtain that microcausality is not satisfied for free 
Dirac field on noncommutative spacetime. However this conclusion is not 
reasonable because it relies on the representation of $\gamma$-matrixes. 
The reason is that for Dirac field, its physical observable quantities 
are not $\overline{\psi}_{\alpha}(x)\star\psi_{\beta}(x)$ directly. 
They are $\overline{\psi}(x)\star\psi(x)$ and 
$\overline{\psi}(x)\gamma^{\mu}\star\psi(x)$ etc. that constructed 
from $\overline{\psi}_{\alpha}(x)\star\psi_{\beta}(x)$ and the 
$\gamma$-matrixes. Therefore we need to analyze the commutators for 
these observable quantities actually.

  For the Lorentz scalar $\overline{\psi}(x)\star\psi(x)$ we need 
to analyze whether 
\begin{eqnarray*}
  B_{0}(x,y) & = & \langle0\vert[:\overline{\psi}(x)\star\psi(x):,
:\overline{\psi}(y)\star\psi(y):]_{\star}\vert0\rangle    \\
& = & \langle0\vert[:\overline{\psi}_{\alpha}(x)\star
      \psi_{\alpha}(x):,:\overline{\psi}_{\sigma}(y)\star
\psi_{\sigma}(y):]_{\star}\vert0\rangle 
\end{eqnarray*}
$$  \eqno{(3.31)}  $$ 
vanishes or not for a spacelike interval of $({\bf x}-{\bf y})$. 
This can be seen from its equal-time commutator 
\begin{eqnarray*}
  B_{0}({\bf x},t,{\bf y},t)
& = & \langle0\vert[:\overline{\psi}({\bf x},t)
\star\psi({\bf x},t):,:\overline{\psi}({\bf y},t)
\star\psi({\bf y},t):]_{\star}\vert0\rangle            \\
& = & \langle0\vert[:\overline{\psi}_{\alpha}({\bf x},t)\star
\psi_{\alpha}({\bf x},t):,:\overline{\psi}_{\sigma}({\bf y},t)
\star\psi_{\sigma}({\bf y},t):]_{\star}\vert0\rangle ~. 
\end{eqnarray*}
$$  \eqno{(3.32)}  $$ 
From Eq. (3.28) we obtain 
\begin{eqnarray*}
  & ~ & \langle0\vert[:\overline{\psi}_{\alpha}({\bf x},t)\star
      \psi_{\alpha}({\bf x},t):,:\overline{\psi}_{\sigma}
       ({\bf y},t)\star\psi_{\sigma}({\bf y},t):]_{\star}
        \vert0\rangle             \\
  & = & \int\frac{d^{3}p_{1}}{(2\pi)^{3}2E_{p_{1}}}
        \int\frac{d^{3}p_{2}}{(2\pi)^{3}2E_{p_{2}}}
         \Big[(mp_{10}\gamma^{0}_{\sigma\alpha}\delta_{\alpha\sigma} 
-mp_{10}\gamma^{0}_{\alpha\sigma}\delta_{\sigma\alpha}    
+mp_{20}\gamma^{0}_{\sigma\alpha}\delta_{\alpha\sigma} 
-mp_{20}\gamma^{0}_{\alpha\sigma}\delta_{\sigma\alpha}  \\
  & ~ & ~~~~~~~~~~~~~~~~~~~~~~~~~~~~~~~~~~~~
+p_{1i}p_{2j}\gamma^{i}_{\sigma\alpha}\gamma^{j}_{\alpha\sigma} 
        -p_{1i}p_{2j}\gamma^{i}_{\alpha\sigma}\gamma^{j}_{\sigma\alpha}) 
\cos({\bf p}_{1}+{\bf p}_{2})\cdot({\bf x}-{\bf y})     \\
  & ~ & 
      -(p_{10}p_{2j}\gamma^{0}_{\sigma\alpha}\gamma^{j}_{\alpha\sigma}
      +p_{1i}p_{20}\gamma^{i}_{\sigma\alpha}\gamma^{0}_{\alpha\sigma}
      +mp_{1i}\gamma^{i}_{\sigma\alpha}\delta_{\alpha\sigma}
-mp_{2j}\delta_{\sigma\alpha}\gamma^{j}_{\alpha\sigma}         \\
  & ~ & ~~
+p_{10}p_{2j}\gamma^{0}_{\alpha\sigma}\gamma^{j}_{\sigma\alpha} 
+p_{1i}p_{20}\gamma^{i}_{\alpha\sigma}\gamma^{0}_{\sigma\alpha} 
      +mp_{1i}\gamma^{i}_{\alpha\sigma}\delta_{\sigma\alpha} 
-mp_{2j}\delta_{\alpha\sigma}\gamma^{j}_{\sigma\alpha})
~ i\sin({\bf p}_{1}+{\bf p}_{2})\cdot({\bf x}-{\bf y})\Big] ~. 
\end{eqnarray*}
$$  \eqno{(3.33)}  $$ 
In Eq. (3.33) for $\alpha$ and $\sigma$ we need to sum up them from 
$1$ to $4$. From the properties of the traces of $\gamma$-matrixes [26], 
we can obtain that the coefficients before  
$\cos({\bf p}_{1}+{\bf p}_{2})\cdot({\bf x}-{\bf y})$ and
$\sin({\bf p}_{1}+{\bf p}_{2})\cdot({\bf x}-{\bf y})$ are all zero 
in Eq. (3.33), and they do not depend on the representation of the 
$\gamma$-matrixes. From the Lorentz invariance of the expression we 
obtain 
$$
B_{0}(x,y)=\langle0\vert[:\overline{\psi}(x)\star\psi(x):,
:\overline{\psi}(y)\star\psi(y):]_{\star}\vert0\rangle=0  
  ~~~~~~ \mbox{for} ~~~~~~ (x-y)^{2}<0  ~. 
  \eqno{(3.34)}  $$
Therefore microcausality is satisfied for the Lorentz scalar 
$\overline{\psi}(x)\star\psi(x)$ of free Dirac field on 
noncommutative spacetime.

  For the current $\overline{\psi}(x)\gamma^{\mu}\star\psi(x)$ of 
the Dirac field, we need to analyze whether 
\begin{eqnarray*}
  B_{0}^{\mu}(x,y) & = & 
 \langle0\vert[:\overline{\psi}(x)\gamma^{\mu}\star\psi(x):, 
:\overline{\psi}(y)\gamma^{\mu}\star\psi(y):]_{\star}
\vert0\rangle    \\
& = & \langle0\vert[:\overline{\psi}_{\alpha}(x)
\gamma^{\mu}_{\alpha\beta}\star\psi_{\beta}(x):,
:\overline{\psi}_{\sigma}(y)\gamma^{\mu}_{\sigma\tau}\star
\psi_{\tau}(y):]_{\star}\vert0\rangle 
\end{eqnarray*}
$$  \eqno{(3.35)}  $$ 
vanishes or not for a spacelike interval of $({\bf x}-{\bf y})$. 
This can be seen from its equal-time commutator 
\begin{eqnarray*}
  B_{0}^{\mu}({\bf x},t,{\bf y},t)
& = & \langle0\vert[:\overline{\psi}({\bf x},t)\gamma^{\mu}
\star\psi({\bf x},t):,:\overline{\psi}({\bf y},t)\gamma^{\mu}
\star\psi({\bf y},t):]_{\star}\vert0\rangle            \\
& = & \langle0\vert[:\overline{\psi}_{\alpha}
({\bf x},t)\gamma^{\mu}_{\alpha\beta}\star
\psi_{\beta}({\bf x},t):,:\overline{\psi}_{\sigma}({\bf y},t)
\gamma^{\mu}_{\sigma\tau}
\star\psi_{\tau}({\bf y},t):]_{\star}\vert0\rangle ~. 
\end{eqnarray*}
$$  \eqno{(3.36)}  $$ 
From Eq. (3.28) we obtain 
\begin{eqnarray*}
  & ~ & \langle0\vert[:\overline{\psi}_{\alpha}
({\bf x},t)\gamma^{\mu}_{\alpha\beta}\star
\psi_{\beta}({\bf x},t):,:\overline{\psi}_{\sigma}({\bf y},t)
\gamma^{\mu}_{\sigma\tau}
\star\psi_{\tau}({\bf y},t):]_{\star}\vert0\rangle             \\
  & = & \int\frac{d^{3}p_{1}}{(2\pi)^{3}2E_{p_{1}}}
        \int\frac{d^{3}p_{2}}{(2\pi)^{3}2E_{p_{2}}}
         \Big[(mp_{10}\gamma^{0}_{\tau\alpha}
\gamma^{\mu}_{\alpha\beta}\delta_{\beta\sigma} 
\gamma^{\mu}_{\sigma\tau}
-mp_{10}\gamma^{0}_{\beta\sigma}\gamma^{\mu}_{\sigma\tau}
\delta_{\tau\alpha}\gamma^{\mu}_{\alpha\beta}    
+mp_{20}\delta_{\beta\sigma}\gamma^{\mu}_{\sigma\tau}
\gamma^{0}_{\tau\alpha}\gamma^{\mu}_{\alpha\beta}       \\
& ~ & -mp_{20}\delta_{\tau\alpha}\gamma^{\mu}_{\alpha\beta} 
       \gamma^{0}_{\beta\sigma}\gamma^{\mu}_{\sigma\tau}
+p_{1i}p_{2j}\gamma^{i}_{\tau\alpha}\gamma^{\mu}_{\alpha\beta}
\gamma^{j}_{\beta\sigma}\gamma^{\mu}_{\sigma\tau}  
-p_{1i}p_{2j}\gamma^{i}_{\beta\sigma}\gamma^{\mu}_{\sigma\tau}
\gamma^{j}_{\tau\alpha}\gamma^{\mu}_{\alpha\beta}) 
\cos({\bf p}_{1}+{\bf p}_{2})\cdot({\bf x}-{\bf y})     \\
  & ~ & 
-(p_{10}p_{2j}\gamma^{0}_{\tau\alpha}\gamma^{\mu}_{\alpha\beta}
\gamma^{j}_{\beta\sigma}\gamma^{\mu}_{\sigma\tau}  
+p_{1i}p_{20}\gamma^{i}_{\tau\alpha}\gamma^{\mu}_{\alpha\beta}
\gamma^{0}_{\beta\sigma}\gamma^{\mu}_{\sigma\tau}
+mp_{1i}\gamma^{i}_{\tau\alpha}\gamma^{\mu}_{\alpha\beta}
\delta_{\beta\sigma}\gamma^{\mu}_{\sigma\tau}
-mp_{2j}\delta_{\tau\alpha}\gamma^{\mu}_{\alpha\beta}
\gamma^{j}_{\beta\sigma}\gamma^{\mu}_{\sigma\tau}     \\
  & ~ & ~~
+p_{10}p_{2j}\gamma^{0}_{\beta\sigma}\gamma^{\mu}_{\sigma\tau}
\gamma^{j}_{\tau\alpha}\gamma^{\mu}_{\alpha\beta}  
 +p_{1i}p_{20}\gamma^{i}_{\beta\sigma}\gamma^{\mu}_{\sigma\tau}
\gamma^{0}_{\tau\alpha}\gamma^{\mu}_{\alpha\beta}  
+mp_{1i}\gamma^{i}_{\beta\sigma}\gamma^{\mu}_{\sigma\tau}
\delta_{\tau\alpha}\gamma^{\mu}_{\alpha\beta} 
-mp_{2j}\delta_{\beta\sigma}\gamma^{\mu}_{\sigma\tau}
\gamma^{j}_{\tau\alpha}\gamma^{\mu}_{\alpha\beta})     \\
  & ~ & ~~~~~
 i\sin({\bf p}_{1}+{\bf p}_{2})\cdot({\bf x}-{\bf y})\Big] ~. 
\end{eqnarray*}
$$  \eqno{(3.37)}  $$ 
In Eq. (3.37), the summations of the indexes are traces of the 
producted $\gamma$-matrixes. From the properties of the traces of 
$\gamma$-matrixes [26], we can obtain that the coefficients before  
$\cos({\bf p}_{1}+{\bf p}_{2})\cdot({\bf x}-{\bf y})$ and
$\sin({\bf p}_{1}+{\bf p}_{2})\cdot({\bf x}-{\bf y})$ are all zero 
in Eq. (3.37), and they do not depend on the representation of the 
$\gamma$-matrixes. From the Lorentz invariance of the expression we 
obtain 
$$
B_{0}^{\mu}(x,y)= 
 \langle0\vert[:\overline{\psi}(x)\gamma^{\mu}\star\psi(x):, 
:\overline{\psi}(y)\gamma^{\mu}\star\psi(y):]_{\star}
\vert0\rangle=0
  ~~~~~~ \mbox{for} ~~~~~~ (x-y)^{2}<0  ~. 
  \eqno{(3.38)}  $$
Therefore microcausality is satisfied for the current  
$\overline{\psi}(x)\gamma^{\mu}\star\psi(x)$ of free Dirac field 
on noncommutative spacetime. For the other bilinear forms of free 
Dirac field on noncommutative spacetime such as 
$\overline{\psi}(x)\gamma^{5}\star\psi(x)$, 
$\overline{\psi}(x)\gamma^{5}\gamma^{\mu}\star\psi(x)$, and 
$\overline{\psi}(x)\sigma^{\mu\nu}\star\psi(x)$, we can also 
obtain that they satisfy the microcausality through explicit 
calculations, and the conclusion does not depend on the 
representation of $\gamma$-matrixes. However for the details 
of the calculations for these bilinear forms, we omit to 
write down them here.

  From the above analysis, we can also see that for the microcausality 
of Dirac field on ordinary commutative spacetime, if we examine it from 
Eq. (3.12), we can obtain that it satisfies the microcausality 
directly. However if we calculate the vacuum expectation values for the 
commutators, we will obtain the result which is equal to Eq. (3.26), 
i.e., 
\begin{eqnarray*}
  & ~ & \langle0\vert[:\overline{\psi}_{\alpha}(x) 
      \psi_{\beta}(x):,:\overline{\psi}_{\sigma}(y) 
      \psi_{\tau}(y):]_{\star}\vert0\rangle             \\
  & = & \int\frac{d^{3}p_{1}}{(2\pi)^{3}2E_{p_{1}}}
        \int\frac{d^{3}p_{2}}{(2\pi)^{3}2E_{p_{2}}}
        \Big[(\not\!{p}_{1}-m)_{\tau\alpha }
(\not\!{p}_{2}+m)_{\beta\sigma}
e^{-i(p_{1}+p_{2})(x-y)}                \\
  & ~ & ~~~~~~~~~~~~~~~~~~~~~~~~~~~~~~~~~~~~~~~
-(\not\!{p}_{1}-m)_{ \beta\sigma}
(\not\!{p}_{2}+m)_{\tau\alpha}
  e^{i(p_{1}+p_{2})(x-y)}\Big] ~.  
\end{eqnarray*}
$$  \eqno{(3.39)}  $$ 
Similarly we will obtain the result as that of Eqs. (3.27)-(3.30). 
Thus from this approach, we cannot obtain that Dirac field satisfies 
the microcausality on ordinary commutative spacetime directly. 
However through the calculation of the expectation values for the 
commutators of $\overline{\psi}(x)\star\psi(x)$ and 
$\overline{\psi}(x)\gamma^{\mu}\star\psi(x)$ etc., we can still 
obtain that Dirac field on ordinary commutative spacetime satisfies 
the microcausality, as that of Eqs. (3.31)-(3.38).

\section{Non-vacuum state expectation values}

\indent 

  For the criterion of microcausality violation given by Eq. (2.5), 
the state vector $\vert\Psi\rangle$ is not a vacuum state generally. 
Therefore we need to calculate the non-vacuum state expectation values 
of $B_{\alpha\beta\sigma\tau}(x,y)$ given by Eq. (3.15). For such a 
purpose we need first to define the state vector $\vert\Psi\rangle$ 
for a Dirac field quantum system. We can write it as   
$$
  \vert\Psi\rangle=\vert N_{p_{1}}(s,s^{\prime})N_{p_{2}}(s,s^{\prime})
         \cdots N_{p_{i}}(s,s^{\prime})\cdots,0\rangle ~.
  \eqno{(4.1)}  $$ 
It is in the occupation eigenstate. 
In Eq. (4.1) we use $N_{p_{i}}$ to represent the occupation number for 
the momentum $p_{i}$, and use $(s,s^{\prime})$ to represent four kinds of 
the spinors $u(p,s)$ and $v(p,s)$. $N_{p_{i}}(s,s^{\prime})$ can only take 
the values $0$ and $1$. We suppose that the occupation numbers are nonzero 
only on some separate momentums $p_{i}$. For all the other momentums, the 
occupation numbers are zero. We use $0$ to represent that the occupation 
numbers are zero on all the other momentums and spins in Eq. (4.1). The 
state vector $\vert\Psi\rangle$ has the following properties [27]: 
$$  
  \langle N_{p_{1}}(s,s^{\prime})N_{p_{2}}(s,s^{\prime})
         \cdots N_{p_{i}}(s,s^{\prime})\cdots\vert
          N_{p_{1}}(s,s^{\prime})N_{p_{2}}(s,s^{\prime})
         \cdots N_{p_{i}}(s,s^{\prime})\cdots\rangle =1 ~,        $$
$$  
  \sum\limits_
      {{\small N_{p_{1}}(s,s^{\prime})N_{p_{2}}(s,s^{\prime})\cdots}}
      \vert N_{p_{1}}(s,s^{\prime})N_{p_{2}}(s,s^{\prime})
         \cdots N_{p_{i}}(s,s^{\prime})\cdots\rangle\langle
     N_{p_{1}}(s,s^{\prime})N_{p_{2}}(s,s^{\prime})
         \cdots N_{p_{i}}(s,s^{\prime})\cdots\vert=1 ~,             $$
$$  \eqno{(4.2)}  $$   
$$
  a_{s,s^{\prime}}(p_{i})\vert N_{p_{1}}(s,s^{\prime})N_{p_{2}}
       (s,s^{\prime})\cdots N_{p_{i-1}}(s,s^{\prime})0_{p_{i}}
       (s,s^{\prime})N_{p_{i+1}}(s,s^{\prime})\cdots\rangle =0 ~,      $$ 
\begin{eqnarray*}
  & ~ & a_{s,s^{\prime}}(p_{i})\vert N_{p_{1}}(s,s^{\prime})N_{p_{2}}
       (s,s^{\prime})\cdots N_{p_{i-1}}(s,s^{\prime})1_{p_{i}}
       (s,s^{\prime})N_{p_{i+1}}(s,s^{\prime})\cdots\rangle      \\ 
  & = &  (-1)^{\sum\limits_{l=1}^{i-1}N_{l}(s,s^{\prime})}     
         \vert N_{p_{1}}(s,s^{\prime})N_{p_{2}}(s,s^{\prime})
         \cdots N_{p_{i-1}}(s,s^{\prime})0_{p_{i}}(s,s^{\prime})
         N_{p_{i+1}}(s,s^{\prime})\cdots\rangle ~,
\end{eqnarray*}  
$$
  a^{\dagger}_{s,s^{\prime}}(p_{i})\vert N_{p_{1}}(s,s^{\prime})
         N_{p_{2}}(s,s^{\prime})
         \cdots N_{p_{i-1}}(s,s^{\prime})1_{p_{i}}(s,s^{\prime})
         N_{p_{i+1}}(s,s^{\prime})\cdots\rangle =0 ~,              $$
\begin{eqnarray*}
  & ~ & a^{\dagger}_{s,s^{\prime}}(p_{i})\vert N_{p_{1}}(s,s^{\prime})
         N_{p_{2}}(s,s^{\prime})
         \cdots N_{p_{i-1}}(s,s^{\prime})0_{p_{i}}(s,s^{\prime})
         N_{p_{i+1}}(s,s^{\prime})\cdots\rangle      \\ 
  & = &  (-1)^{\sum\limits_{l=1}^{i-1}N_{l}(s,s^{\prime})}     
         \vert N_{p_{1}}(s,s^{\prime})N_{p_{2}}(s,s^{\prime})
         \cdots N_{p_{i-1}}(s,s^{\prime})1_{p_{i}}(s,s^{\prime})
         N_{p_{i+1}}(s,s^{\prime})\cdots\rangle ~.
\end{eqnarray*}  
$$  \eqno{(4.3)}  $$ 
In Eq. (4.3), we use $a_{s,s^{\prime}}$ to represent one kind of the  
annihilation operators $b(p,s)$ and $d(p,s)$, and use 
$a^{\dagger}_{s,s^{\prime}}$ to represent one kind of the creation 
operators $b^{\dagger}(p,s)$ and $d^{\dagger}(p,s)$. We need to 
calculate the function 
$$
  B_{\alpha\beta\sigma\tau}(x,y)
      =\langle\Psi\vert[:\overline{\psi}_{\alpha}(x)\star
      \psi_{\beta}(x):,:\overline{\psi}_{\sigma}(y)\star
      \psi_{\tau}(y):]_{\star}\vert\Psi\rangle ~. 
  \eqno{(4.4)}  $$

  In the scalar field case [23], we have proved that the non-vacuum 
state expectation value for the Moyal commutator 
$[:\varphi(x)\star\varphi(x):,:\varphi(y)\star\varphi(y):]_{\star}$ 
is just equal to the corresponding vacuum expectation value. It is a 
universal function for an arbitrary state vector of scalar field 
quantum system. 
We can also prove that such a property also hold for Dirac field. 
We first need to recognize that the total energy of an actual field 
quantum system is always finite. This makes the occupation numbers 
$N_{p_{i}}(s,s^{\prime})$ be nonzero only on a set of finite number 
separate momentums $p_{i}$, because $N_{p_{i}}(s,s^{\prime})$ take 
values of the integral numbers $0$ and $1$. If $N_{p_{i}}$ take nonzero 
values on infinite number separate momentums $p_{i}$ or on a continuous 
interval of the momentum, the total energy of the field quantum system 
will be infinite.

  Through calculation we can write
$$
  \langle\Psi\vert[:\overline{\psi}_{\alpha}(x)\star
      \psi_{\beta}(x):,:\overline{\psi}_{\sigma}(y)\star
      \psi_{\tau}(y):]_{\star}\vert\Psi\rangle = 
\int\frac{d^{3}p_{1}}{(2\pi)^{3}2E_{p_{1}}} 
        \int\frac{d^{3}p_{2}}{(2\pi)^{3}2E_{p_{2}}}
       G_{\alpha\beta,\sigma\tau}(p_{1},p_{2},x,y) ~.  
  \eqno{(4.5)}  $$ 
From Eq. (3.23) we can see that the non-zero contributions to the 
integrand $G_{\alpha\beta,\sigma\tau}(p_{1},p_{2},x,y)$ not only 
come from the operators 
$\overline{\psi}_{\alpha}^{+}(x)\star\psi_{\beta}^{+}(x)
\star\overline{\psi}_{\sigma}^{-}(y)\star\psi_{\tau}^{-}(y)$ and 
$\overline{\psi}_{\sigma}^{+}(y)\star\psi_{\tau}^{+}(y)
  \star\overline{\psi}_{\alpha}^{-}(x)\star\psi_{\beta}^{-}(x)$ 
as that of Eq. (3.26), but also come from the operators 
$\psi_{\beta}^{-}(x)\star\overline{\psi}_{\alpha}^{+}(x)\star 
         \psi_{\tau}^{-}(y)\star\overline{\psi}_{\sigma}^{+}(y)$, 
$\psi_{\beta}^{-}(x)\star\overline{\psi}_{\alpha}^{+}(x)\star
         \overline{\psi}_{\sigma}^{-}(y)\star\psi_{\tau}^{+}(y)$, 
$\overline{\psi}_{\alpha}^{-}(x)\star\psi_{\beta}^{+}(x)\star 
         \psi_{\tau}^{-}(y)\star\overline{\psi}_{\sigma}^{+}(y)$,    
$\overline{\psi}_{\alpha}^{-}(x)\star\psi_{\beta}^{+}(x)\star
         \overline{\psi}_{\sigma}^{-}(y)\star\psi_{\tau}^{+}(y)$, 
$\overline{\psi}_{\alpha}^{-}(x)\star\psi_{\beta}^{-}(x)\star
         \overline{\psi}_{\sigma}^{+}(y)\star\psi_{\tau}^{+}(y)$ 
and 
$\psi_{\tau}^{-}(y)\star\overline{\psi}_{\sigma}^{+}(y)\star 
         \psi_{\beta}^{-}(x)\star\overline{\psi}_{\alpha}^{+}(x)$, 
$\overline{\psi}_{\sigma}^{-}(y)\star\psi_{\tau}^{+}(y)\star
\psi_{\beta}^{-}(x)\star\overline{\psi}_{\alpha}^{+}(x)$, 
$\psi_{\tau}^{-}(y)\star\overline{\psi}_{\sigma}^{+}(y)\star
\overline{\psi}_{\alpha}^{-}(x)\star\psi_{\beta}^{+}(x)$,    
$\overline{\psi}_{\sigma}^{-}(y)\star\psi_{\tau}^{+}(y)\star
\overline{\psi}_{\alpha}^{-}(x)\star\psi_{\beta}^{+}(x)$, 
$\overline{\psi}_{\sigma}^{-}(y)\star\psi_{\tau}^{-}(y)\star
\overline{\psi}_{\alpha}^{+}(x)\star\psi_{\beta}^{+}(x)$. 
These operators all have the equal numbers of negative frequency and 
positive frequency components. This makes the integrand 
$G_{\alpha\beta,\sigma\tau}(p_{1},p_{2},x,y)$ of Eq. (4.5) be not equal 
to the integrand of Eq. (3.26) generally. However because in the state 
vector $\vert\Psi\rangle$, the occupation numbers 
$N_{p_{i}}(s,s^{\prime})$ are nonzero only on a set of finite number 
separate momentums $p_{i}$ as pointed out above, the integrand 
$G_{\alpha\beta,\sigma\tau}(p_{1},p_{2},x,y)$ of Eq. (4.5) only changes 
its value relative to the integrand of Eq. (3.26) on a set of finite 
number separate momentums $p_{i}$ of $\vert\Psi\rangle$, just like that 
of the scalar field case. However we omit to write down the detailed 
analysis for such a fact here. Thus we need not to obtain the explicit 
form for the integrand $G_{\alpha\beta,\sigma\tau}(p_{1},p_{2},x,y)$ 
of Eq. (4.5) on the set of finite number separate momentums $p_{i}$ of 
$\vert\Psi\rangle$. This is because the integrand 
$G_{\alpha\beta,\sigma\tau}(p_{1},p_{2},x,y)$ may be a bounded 
function. The total integral measure for a set of finite number separate 
momentums $p_{i}$ is zero. Therefore according to the theory of 
integration (for example see Ref. [28]), the total integral of Eq. (4.5) 
is not changed from that of Eq. (3.26). Thus we obtain 
\begin{eqnarray*}
  & ~ & \langle\Psi\vert[:\overline{\psi}_{\alpha}(x)\star
      \psi_{\beta}(x):,:\overline{\psi}_{\sigma}(y)\star
      \psi_{\tau}(y):]_{\star}\vert\Psi\rangle           \\
  & = &   \int\frac{d^{3}p_{1}}{(2\pi)^{3}2E_{p_{1}}}
        \int\frac{d^{3}p_{2}}{(2\pi)^{3}2E_{p_{2}}}
        \Big[(\not\!{p}_{1}-m)_{\tau\alpha }
(\not\!{p}_{2}+m)_{\beta\sigma}
e^{-i(p_{1}+p_{2})(x-y)}                \\
  & ~ & ~~~~~~~~~~~~~~~~~~~~~~~~~~~~~~~~~~~~~~~
-(\not\!{p}_{1}-m)_{ \beta\sigma}
(\not\!{p}_{2}+m)_{\tau\alpha}
           e^{i(p_{1}+p_{2})(x-y)}\Big] ~.  
\end{eqnarray*}
$$  \eqno{(4.6)}  $$ 
Or we can write 
$$
  B_{\alpha\beta\sigma\tau}(x,y)=B_{0,\alpha\beta\sigma\tau}(x,y) ~,
  \eqno{(4.7)}  $$
which is a universal function for an arbitrary state vector of 
Eq. (4.1). Thus from the result of Sec. IV we can derive 
$$
  \langle\Psi\vert[:\overline{\psi}(x)\star\psi(x):,
:\overline{\psi}(y)\star\psi (y):]_{\star}\vert\Psi\rangle=0 
  ~~~~~~~~ \mbox{for} ~~~~~~~~ (x-y)^{2}<0 ~, 
  \eqno{(4.8)}  $$ 
and similarly 
$$
  \langle\Psi\vert[:\overline{\psi}(x)\gamma^{\mu}\star\psi(x):,
:\overline{\psi}(y)\gamma^{\mu}\star\psi (y):]
_{\star}\vert\Psi\rangle=0 
  ~~~~~~ \mbox{for} ~~~~~~ (x-y)^{2}<0 ~. 
  \eqno{(4.9)}  $$ 
For the other bilinear forms of the Dirac field such as 
$\overline{\psi}(x)\gamma^{5}\star\psi(x)$, 
$\overline{\psi}(x)\gamma^{5}\gamma^{\mu}\star\psi(x)$, and 
$\overline{\psi}(x)\sigma^{\mu\nu}\star\psi(x)$ on noncommutative 
spacetime, we can obtain the same result. This means that microcausality 
is satisfied for free Dirac field on noncommutative spacetime generally, 
no matter whether $\theta^{0i}=0$ or $\theta^{0i}\neq0$.

\section{Conclusion}

\indent

  In this paper, we studied the microcausality of free Dirac field on 
noncommutative spacetime. As pointed out in the Introduction, we expand 
quantum fields on noncommutative spacetime according to their usual 
Lorentz invariant spectral measures. For the $SO(1,1)\times SO(2)$ 
invariant spectral measures as that constructed in Ref. [10], we would 
rather consider that it is an effective result evoked from the nonlocal 
interactions of NCFTs as pointed out in Ref. [20]. Therefore we consider 
that the infinite propagation speed of the waves inside the $SO(1,1)$ 
light wedge is in fact an effect of the nonlocal interactions of NCFTs, 
while not the property of free fields. Similarly, we would rather 
consider that the twisted Poincar{\' e} invariance [13] of NCFTs is also 
an effective property that generated from the nonlocal interactions of 
NCFTs. For NCFTs, although there exist the nonlocality, it does not mean 
that Lorentz invariance is necessarily broken. In fact, the nonlocality 
of NCFTs can be in self-consistency with the Lorentz invariance. For 
the noncommutative parameters $\theta^{\mu\nu}$, if we take them to 
be a second-order antisymmetric tensor, it is possible for us to 
realize the Lorentz invariance of NCFTs [15-18]. On the other hand 
we know that the breakdown of Lorentz 
invariance has not been discovered yet in the experiments [19]. It is 
reasonable to believe that Lorentz invariance is a more fundamental 
principle of physics in a local area, although it may not be satisfied 
for a large scale structure of the universe. Therefore we expand 
quantum fields on noncommutative spacetime according to their usual 
Lorentz invariant spectral measures, except that we take the 
coordinates to be noncommutative operators.

  Another reason for us to do so is that in 
most occasions for the perturbative calculations of NCFTs in the 
literature, the propagators of quantum fields on noncommutative spacetime 
are obtained based on the usual Lorentz invariant spectral measures for 
the expansion of fields. Thus it is necessary for us to study whether 
microcausality is violated or not for quantum fields on noncommutative 
spacetime if we expand them according to their usual Lorentz invariant 
spectral measures, i.e., we hope to study whether the breakdown of 
microcausality can occur in NCFTs in accordance with the Lorentz 
invariance. In Ref. [22], Greenberg obtained that microcausality is 
violated for free scalar field 
on noncommutative spacetime generally no matter whether $\theta^{0i}$ 
vanishes or not. However we have pointed out in Ref. [23] that there 
are some problems for the arguments in Ref. [22] for the result of the 
microcausality violation. We obtained that microcausality is violated 
for the quadratic operators of scalar field on noncommutative spacetime 
only when $\theta^{0i}\neq0$ [23].

  For the free Dirac field on ordinary commutative spacetime, we can 
obtain that microcausality is satisfied directly from Eqs. (3.12) and 
(3.13) according to the properties of the fundamental anticommutators 
of Dirac field. However for the Dirac field on noncommutative 
spacetime, because its fundamental Moyal anticommutators are not the 
$c$-number functions [24], we cannot obtain the conclusion that 
microcausality is satisfied for Dirac field on noncommutative spacetime 
from Eq. (3.14). We need to calculate the vacuum and non-vacuum state 
expectation values for the Moyal commutator 
$[\overline{\psi}_{\alpha}(x)\star\psi_{\beta}(x),\overline{\psi}_
{\sigma}(x^{\prime})\star\psi_{\tau}(x^{\prime})]_{\star}$. 
As demonstrated in Sec. II, this is also the demand of the physical 
measurements. In Sec. IV, we argued that because the total energy of 
an actual field quantum system is always finite, the non-vacuum state 
expectation value for the Moyal commutator 
$[\overline{\psi}_{\alpha}(x)\star\psi_{\beta}(x),\overline{\psi}_
{\sigma}(x^{\prime})\star\psi_{\tau}(x^{\prime})]_{\star}$ is just 
equal to its vacuum expectation value. It is a universal function for 
an arbitrary state vector of the Dirac field quantum system as that 
of the scalar field case [23].

  In Sec. III, we obtained that microcausality is not satisfied 
generally for the Moyal commutator 
$[\overline{\psi}_{\alpha}(x)\star\psi_{\beta}(x),\overline{\psi}_
{\sigma}(x^{\prime})\star\psi_{\tau}(x^{\prime})]_{\star}$ 
for some cases of the indexes $\alpha$, $\beta$, $\sigma$, and $\tau$, 
and the result depends on the choice of the representation of 
$\gamma$-matrixes, as that given by Eqs. (3.29) and (3.30). However, 
for the physical observables of Dirac field on noncommutative 
spacetime such as the Lorentz scalar $\overline{\psi}(x)\star\psi(x)$ 
and the current $\overline{\psi}(x)\gamma^{\mu}\star\psi(x)$, we have  
obtained that microcausality is satisfied. For some other bilinear forms 
of the Dirac field such as $\overline{\psi}(x)\gamma^{5}\star\psi(x)$, 
$\overline{\psi}(x)\gamma^{5}\gamma^{\mu}\star\psi(x)$, and 
$\overline{\psi}(x)\sigma^{\mu\nu}\star\psi(x)$, we can also obtain 
that microcausality is satisfied for them on noncommutative spacetime. 
Therefore microcausality is not violated for the free Dirac on 
noncommutative spacetime generally. We have not found the violation of 
microcausality of free Dirac on noncommutative spacetime in accordance 
with the Lorentz invariance, no matter whether $\theta^{0i}$ vanishes 
or not. Therefore there does not exist the information and interaction 
with the transmit speed faster than the speed of light for the free 
Dirac on noncommutative spacetime.

  For the free electromagnetic field on noncommutative spacetime, we 
can expect that its microcausality property is similar to the free 
scalar field case, i.e., there may exist the violation of microcausality 
for the free electromagnetic field on noncommutative spacetime when 
$\theta^{0i}\neq0$. As pointed out in Refs. [29,30], unitarity of the 
$S$-matrixes may be lost for NCFTs when $\theta^{0i}\neq0$. Therefore 
if we exclude the case of $\theta^{0i}\neq0$ for the noncommutative 
parameters $\theta^{\mu\nu}$ from the demand of the unitarity of the 
$S$-matrixes, microcausality is also satisfied for free bose fields 
on noncommutative spacetime. However one can be conscious of that 
it is rather difficult for us to consider that the time coordinate is 
in an unequal position to the space coordinates even if the spacetime 
coordinates are noncommutative. For the unitarity problem of NCFTs 
with $\theta^{0i}\neq0$, some authors have devoted in searching 
for the retrieving possibilities [31-33].

  Although for the free fields on noncommutative spacetime there may 
not exist the violation of microcausality generally, and therefore 
there does not exist the information and interaction with the transmit 
speed faster than the speed of light for the free fields on 
noncommutative spacetime, except for the free bose fields for the 
spacetime noncommutativity with $\theta^{0i}\neq0$, there can exist the 
infinite propagation speed for the signals and interactions on 
noncommutative spacetime as that indicated in Refs. [6-8], no matter 
whether $\theta^{0i}$ vanishes or not. On the other hand, the nonlocal 
interactions of NCFTs can result the existence of lower dimensional 
light wedge [20]. This can result the effective spectral measures for 
the expansions of quantum fields on noncommutative spacetime in the 
form of the light wedge invariance as that constructed in Ref. [10] 
and therefore results the infinite propagation speed inside the 
light wedge.

\vskip 1.2cm

\centerline{\bf ACKNOWLEDGEMENTS} 

\vskip 0.2cm

\indent 

  Thanks very much for C.S. Chu, D.H.T. Franco, and S. Pasquetti 
to inform me Refs. [20], [12], and [33].

\vskip 1.2cm

{\bf APPENDIX: SELF-CONSISTENCY OF SPACETIME COMMUTATION 
               RELATIONS (1.3) WITH (1.1)}

\vskip 0.3cm

\indent 

  In this Appendix, we give a proof for the self-consistency of Eq. (1.3) 
with Eq. (1.1) for the spacetime commutation relations. First we have the 
commutation relations (1.1) for the coordinates at the same spacetime 
point
$$
  [x^{\mu},x^{\nu}]=i\theta^{\mu\nu} ~. 
  \eqno{({\rm A}1)}  $$
For a different spacetime point $y$ we also have 
$$
  [y^{\mu},y^{\nu}]=i\theta^{\mu\nu} ~. 
  \eqno{({\rm A}2)}  $$
If we generalize the commutation relations (1.1) to two different 
spacetime points, then we have 
$$
  [x^{\mu},y^{\nu}]=i\theta^{\mu\nu} ~. 
  \eqno{({\rm A}3)}  $$
At the same time equivalently we have 
$$
  [y^{\mu},x^{\nu}]=i\theta^{\mu\nu} ~. 
  \eqno{({\rm A}4)}  $$
To introduce the difference $\Delta x^{\mu}$ between $x^{\mu}$ and 
$y^{\mu}$, we have 
$$
  x^{\mu}+\Delta x^{\mu}=y^{\mu} ~.
  \eqno{({\rm A}5)}  $$
Similarly to introduce the difference $\Delta x^{\nu}$ between $x^{\nu}$ 
and $y^{\nu}$, we have 
$$
  x^{\nu}+\Delta x^{\nu}=y^{\nu} ~.
  \eqno{({\rm A}6)}  $$
From Eqs. ({\rm A}3) and ({\rm A}6) we have 
$$
  [x^{\mu},x^{\nu}+\Delta x^{\nu}]=i\theta^{\mu\nu} ~.  
  \eqno{({\rm A}7)}  $$
To combine Eq. ({\rm A}1) we obtain from Eq. ({\rm A}7)
$$
  [x^{\mu},\Delta x^{\nu}]=0 ~.  
  \eqno{({\rm A}8)}  $$
Similarly, from Eqs. ({\rm A}4) and ({\rm A}5) we have 
$$
  [x^{\mu}+\Delta x^{\mu},x^{\nu}]=i\theta^{\mu\nu} ~.  
  \eqno{({\rm A}9)}  $$
To combine Eq. ({\rm A}1) we obtain from Eq. ({\rm A}9)
$$
  [\Delta x^{\mu},x^{\nu}]=0 ~.  
  \eqno{({\rm A}10)}  $$
To use Eqs. ({\rm A}5) and ({\rm A}6), we can write Eq. ({\rm A}2) as
$$
  [x^{\mu}+\Delta x^{\mu},x^{\nu}+\Delta x^{\nu}]=i\theta^{\mu\nu} ~.  
  \eqno{({\rm A}11)}  $$
To combine Eqs. ({\rm A}1), ({\rm A}8), and ({\rm A}10), we obtain from 
Eq. ({\rm A}.11) that the relation 
$$
  [\Delta x^{\mu},\Delta x^{\nu}]=0   
  \eqno{({\rm A}12)}  $$
should be satisfied, if the relations ({\rm A}3) or equivalently 
({\rm A}4) are in consistence with the relations ({\rm A}1) or 
equivalently ({\rm A}2). We can verify this equation.

  To use Eqs. ({\rm A}5) and ({\rm A}6), we can write 
$$
  [\Delta x^{\mu},\Delta x^{\nu}]=[y^{\mu}-x^{\mu},y^{\nu}-x^{\nu}] ~.
  \eqno{({\rm A}13)}  $$
To expand the right hand side of Eq. ({\rm A}13), we have 
$$
  [y^{\mu}-x^{\mu},y^{\nu}-x^{\nu}]=[y^{\mu},y^{\nu}]-[y^{\mu},x^{\nu}]
          -[x^{\mu},y^{\nu}]+[x^{\mu},x^{\nu}] ~.
  \eqno{({\rm A}14)}  $$
To use Eqs. ({\rm A}1)-({\rm A}4), we obtain 
$$
  [y^{\mu}-x^{\mu},y^{\nu}-x^{\nu}]=0 ~.
  \eqno{({\rm A}15)}  $$
Thus Eq. ({\rm A}12) is satisfied. This means that the relations 
({\rm A}3) or ({\rm A}4) are in consistence with the relations 
({\rm A}1) or ({\rm A}2). Thus the generalization of the spacetime 
commutative relations (1.1) defined at the same spacetime point to 
the spacetime commutative relations (1.3) defined on two different 
spacetime points is reasonable. Or we can write Eqs. ({\rm A}3) 
and ({\rm A}4) in the form of Eq. (1.3) 
$$
  [x_{1}^{\mu},x_{2}^{\nu}]=i\theta^{\mu\nu} ~. 
  \eqno{({\rm A}16)}  $$
We need to take notice that the origin of coordinates in noncommutative 
spacetime is not a $c$-number either. It is also an operator. We can 
write it in the form of boldface as ${\bf 0}$. We have 
$$
  [x^{\mu},{\bf 0}^{\nu}]=i\theta^{\mu\nu} ~,  
  \eqno{({\rm A}17)}  $$
and similarly 
$$
  [{\bf 0}^{\mu},x^{\nu}]=i\theta^{\mu\nu} ~.  
  \eqno{({\rm A}18)}  $$
Thus we need to notice that 
$x^{\mu}-{\bf 0}^{\mu}=\Delta x^{\mu}\neq x^{\mu}$.

\vskip 1.2cm 

\noindent {\large {\bf References}}

\vskip 20pt

[1] H.S. Snyder, Phys. Rev. {\bf 71}, 38 (1947). 

[2] S. Doplicher, K. Fredenhagen, and J.E. Roberts, Phys. Lett. B 
    {\bf 331}, 39 (1994); 

    ~~~ Commun. Math. Phys. {\bf 172}, 187 (1995), hep-th/0303037.

[3] N. Seiberg and E. Witten, J. High Energy Phys. 09 (1999) 032, 
hep-th/9908142. 

[4] M.R. Douglas and N.A. Nekrasov, Rev. Mod. Phys. {\bf 73}, 977 
(2001), hep-th/0106048. 

[5] R.J. Szabo, Phys. Rep. {\bf 378}, 207 (2003), hep-th/0109162. 

[6] M. Van Raamsdonk, J. High Energy Phys. 11 (2001) 006, hep-th/0110093.  

[7] A. Hashimoto and N. Itzhaki, Phys. Rev. D {\bf 63}, 126004 (2001), 
hep-th/0012093. 

[8] B. Durhuus and T. Jonsson, J. High Energy Phys. 10 (2004) 050, 
hep-th/0408190. 

[9] S. Minwalla, M. Van Raamsdonk, and N. Seiberg, J. High Energy Phys. 
    02 (2000)

    ~~~ 020, hep-th/9912072.  

[10] L. {\' A}lvarez-Gaum{\' e} and M.A. V{\' a}zquez-Mozo, 
Nucl. Phys. {\bf B668}, 293 (2003), 

~~~~~ hep-th/0305093. 
  
[11] L. Alvarez-Gaum{\' e}, J.L.F. Barb{\' o}n, and R. Zwicky, 
    J. High Energy Phys. 05 (2001) 

    ~~~~~ 057, hep-th/0103069.

[12] D.H.T. Franco and C.M.M. Polito, J. Math. Phys. {\bf 46}, 
083503 (2005), hep-th/0403028. 

[13] M. Chaichian, P. Pre{\v s}najder, and A. Tureanu, Phys. Rev. Lett. 
{\bf 94}, 151602 (2005), 

  ~~~~~  hep-th/0409096. 

[14] M. Chaichian, K. Nishijima, and A. Tureanu, Phys. Lett. B 
{\bf 633}, 129 (2006), 

  ~~~~~ hep-th/0511094. 

[15] C.E. Carlson, C.D. Carone, and N. Zobin, Phys. Rev. D {\bf 66}, 
075001 (2002), 

  ~~~~~ hep-th/0206035. 

[16] H. Kase, K. Morita, Y. Okumura, and E. Umezawa, Prog. Theor. Phys. 
{\bf 109}, 663 

     ~~~~~ (2003), hep-th/0212176. 

[17] K. Morita, Y. Okumura, and E. Umezawa, Prog. Theor. Phys. 
{\bf 110}, 989 (2003), 

~~~~~ hep-th/0309155.  

[18] R. Banerjee, B. Chakraborty, and K. Kumar, Phys. Rev. D {\bf 70},  
125004 (2004), 

     ~~~~~ hep-th/0408197. 

[19] A. Anisimov, T. Banks, M. Dine, and M. Graesser, 
Phys. Rev. D {\bf 65}, 085032 (2002), 

~~~~~ hep-ph/0106356. 

[20] C.S. Chu, K. Furuta, and T. Inami, Int. J. Mod. Phys. A {\bf 21}, 
 67 (2006), 

~~~~~ hep-th/0502012. 

[21] M. Chaichian, K. Nishijima, and A. Tureanu, Phys. Lett. B 
{\bf 568}, 146 (2003), 

~~~~~ hep-th/0209008. 

[22] O.W. Greenberg, Phys. Rev. D {\bf 73}, 045014 (2006), 
hep-th/0508057. 

[23] Z.Z. Ma, hep-th/0603054. 

[24] Z.Z. Ma, hep-th/0601094. 

[25] J.D. Bjorken and S.D. Drell, {\sl Relativistic quantum fields}   
     (McGraw-Hill, 1965). 

[26] C. Itzykson and J.-B. Zuber, {\sl Quantum field theory}   
     (McGraw-Hill Inc., 1980). 

[27] L.D. Landau and E.M. Lifshitz, {\sl Quantum mechanics} 
     (Pergamon Press, 1977).  

[28] G. de Barra, {\sl Measure theory and integration} 
(Halsled Press, New York, 1981). 

[29] J. Gomis and T. Mehen, Nucl. Phys. {\bf B591}, 265 (2000), 
hep-th/0005129. 

[30] A. Bassetto, F. Vian, L. Griguolo, and G. Nardelli, J. High 
     Energy Phys. 07 (2001) 

     ~~~~~ 008, hep-th/0105257. 

[31] D. Bahns, S. Doplicher, K. Fredenhagen, and G. Piacitelli, 
Phys. Lett. B {\bf 533}, 178  

     ~~~~~ (2002), hep-th/0201222. 

[32] C.S. Chu, J. Lukierski, and W.J. Zakrzewski, Nucl. Phys.  
{\bf B632}, 219 (2002), 

     ~~~~~ hep-th/0201144. 

[33] N. Caporaso and S. Pasquetti, hep-th/0511127.

\end{document}